\documentclass[]{article}

\usepackage{amsmath}
\usepackage{bm}
\usepackage[dvips,xdvi]{graphicx}
\newcommand{\citen}[1]{\cite{#1}}

\begin{document}

%%%%
%\setlength{\baselineskip}{7mm}
%%%%

%------------------------------------------------------------------------------
% title

\title{\bf Density Functional Simulation of Spontaneous Formation of Vesicle in Block Copolymer Solutions}
\author{Takashi Uneyama \\
\\
Department of Physics, Graduate School of Science, Kyoto University \\
Sakyo-ku, Kyoto 606-8502, JAPAN
}
\date{}

\maketitle

%%%%
%\newpage
%%%%

%------------------------------------------------------------------------------
% abstract
\begin{abstract}

%%%%
%\setlength{\baselineskip}{7mm}
%%%%

We carry out numerical simulations of vesicle formation
based on the density functional theory for block copolymer solutions.
It is shown by solving the time evolution equations for concentrations
that a polymer vesicle is spontaneously formed from
the homogeneous state. The vesicle formation mechanism obtained by our
simulation agree with the results of other simulations based on the
particle models as well as experiments.
By changing parameters such as the volume fraction of polymers or the Flory-Huggins
interaction parameter between the hydrophobic subchains and solvents,
we can obtain the spherical micelles, cylindrical micelles or
bilayer structures, too. We also show that the morphological transition
dynamics of the micellar structures can be reproduced by controlling the
Flory-Huggins interaction parameter.
\end{abstract}

%\clearpage

%------------------------------------------------------------------------------
% main text

\section{Introduction}
\label{introduction}
Amphiphilic block copolymers, which consists of hydrophilic subchains and
hydrophobic subchains, are known to form micellar structures, such as
spherical micelles, cylindrical micelles, and vesicles
\cite{Disher-Eisenberg-2002,Choucair-Eisenberg-2003}.
To clarify how these structures are self-organised is a basic problem of
the kinetics of micellar systems.

Several computer simulations have been available to study formation of a
vesicle based on the particle models
(Brownian dynamics (BD) \cite{Noguchi-Takasu-2001}, dissipative particle
dynamics (DPD) \cite{Yamamoto-Maruyama-Hyodo-2002}, and molecular
dynamics (MD) \cite{Marrink-Mark-2003}).
The spontaneous vesicle formation process from the
homogeneous state
observed in these simulations is as follows: The amphiphilic block
copolymers aggregate into small spherical micelles rapidly from the
homogeneous initial state. The spherical
micelles grows to larger micelles by collision (cylindrical micelles, open
disk-like micelles). The large disk-like micelles finally close up and
form vesicles. The micelle growth process and the closure
process are slower than the first spherical micelle formation process.
Hereafter we call this process the ``mechanism I'' (see also Figure
\ref{vesicle_formation_mechanisms}(a)). Note that the mechanism I is
also supported by Monte Carlo simulations \cite{Bernardes-1996} or
experiments of lipid systems
\cite{Leng-Egelhaaf-Cates-2002,Leng-Egelhaaf-Cates-2003}.
Strictly speaking, the mechanism obtained by the experiments for lipid
systems may be different from one for polymer systems. However, since the DPD
simulations for polymer systems and the BD and MD simulations for lipid
systems give similar results (mechanism I), we believe that the vesicle
formation mechanism is common for polymer systems and lipid systems.
We also note that the similar mechanism obtained for polymer systems
by the experiments of morphological transition of cylindrical micelles
to vesicles \cite{Chen-Shen-Eisenberg-1999}.

The field theoretical approach, which has been developed to study mesoscale structures
of block copolymers \cite{Leibler-1980,Ohta-Kawasaki-1986,BohbotRaviv-Wang-2000,Matsen-Schick-1994,Matsen-Bates-1996,Fraaije-1993,Fredrickson-Ganesan-Drolet-2002},
is considered to be useful for vesicle formation, but most of the works
have been limited to simulations in thermal equilibrium
\cite{He-Liang-Huang-Pan-2004,Uneyama-Doi-2005a}.
Since these simulations ignore realistic kinetics (for example, local
mass conservation is not satisfied), the vesicle formation process observed in these
simulations are different from the mechanism I:
The first process is similar to the mechanism I. Small spherical
micelles are formed rapidly. The spherical micelles then grow up to
large spherical micelles by the evaporation-condensation like process.
The large spherical micelles are energetically
unfavourable, and thus the large micelles take the solvents into them to
lower the energy. We call this process the
``mechanism II'' (see also Figure \ref{vesicle_formation_mechanisms}(b)).

The dynamical simulations for diblock copolymer solutions were carried by
Sevink and Zvelindovsky \cite{Sevink-Zvelindovsky-2005}, using the
dynamic self consistent field (SCF) theory. However, in their simulations both subchains are
hydrophobic and the resulting structures are so-called onion structures
\cite{Koizumi-Hasegawa-Hashimoto-1994,Ohta-Ito-1995,Uneyama-Doi-2005,Sevink-Zvelindovsky-2005}.
The onion structures obtained
by simulations are essentially microphase separation structure in the block
copolymer rich droplets in the solvents, and qualitatively different
from the multilayer
vesicle structures which is often called as ``onion structures'' in surfactant
solutions. The vesicle structures contain solvents
inside them and therefore different from these polymer onion structures,
and one should distinguish the onion formation process from the vesicle
formation process. We also note that the simulations for onion structures were
carried out for weak segregation region and this does not agree with
previous equilibrium simulations for vesicles, since
vesicles are observed in rather strong
segregation region. (In this work, we use the word ``strong
segregation region'' as the region where the minimum and maximum values
of the density field is approximately $0$ and $1$. One may call such region as
the intermediate segregation region.)
It should be noted that the formation processes of the onion
structures by simulations \cite{Ohta-Ito-1995,Sevink-Zvelindovsky-2005}
are mainly the lamellar ordering process in droplet like regions.
Therefore the resulting morphologies (onions) are similar to the phase
separation patterns in droplets \cite{Fraaije-Sevink-2003} than vesicles.
Thus we consider that the onion formation process is not the mechanism I.

Recently, dynamical simulations of vesicle formation
have been done by
He and Schmid \cite{He-Schmid-2006}, using the external potential
dynamics (EPD) \cite{Maurits-Fraaije-1997}. They obtained vesicles, but
their formation process is similar to the mechanism II. Unfortunately
the mechanism II is
qualitatively different from the mechanism I, and this means that their
result does not agree with particle simulations or experiments
(we will discuss the reason for this difference in Section
\ref{comparison_with_epd_simulations}).
We expect that the mechanisms observed by different simulation methods
should be the same.

In the present work, we apply our previous model, the density functional theory
for block copolymers
\cite{Uneyama-Doi-2005,Uneyama-Doi-2005a},
to the dynamics. We carry out numerical simulations for
 amphiphilic diblock copolymer solutions in three dimensions.
By using the continuous field model, we show, for the first time,
that a vesicle is spontaneously formed from a disordered uniform phase.
The simulation result is consistent with the mechanism I.
We also show that we can simulate the spontaneous formation process of
various micellar structures (such as spherical micelles or cylindrical
micelles) and the morphological transition dynamics.

%------------------------------------------------------------------------------
\section{Theory}
\label{theory}
The dynamics of block copolymer systems are well studied by using
the dynamic SCF theory \cite{Fraaije-1993}
as well as
the time-dependent Ginzburg-Landau (TDGL) equation \cite{Bahiana-Oono-1990}.
The dynamic SCF simulations are known to be accurate for from weak
segregation region to strong segregation region, but they consume
memory and need large CPU power. In contrast, the TDGL simulations need
less CPU power and enables large scale simulations.
However, the free energy functionals
\cite{Leibler-1980,Ohta-Kawasaki-1986} used in the TGDL approach
is generally not appropriate for the strong segregation region, since
the validity of the Ginzburg-Landau (GL) expansion is guaranteed only for the
weak segregation region where the density fluctuation
is sufficiently small \cite{Kawakatsu-book,Matsen-Bates-1996,Uneyama-Doi-2005}.
Actually Maurits and Fraaije \cite{Maurits-Fraaije-1997b} showed that
the widely used fourth-order
GL expansion model is not sufficient for dynamical
simulations.
To overcome this limitation we need
to use free energy functional model such as the Flory-Huggins-de
Gennes-Lifshitz free energy
\cite{deGennes-1980,Grosberg-Khokhlov-book,Frusawa-2005}, which is not
expressed in the GL expansion form.

In previous simulations, vesicles are observed for rather strong
segregation region in diblock copolymer solutions. Thus we can conclude
that the use of
inappropriate free energy functional
models qualitatively affects micellar structures in 
diblock copolymer solutions. We need to use a free energy
model and a dynamic equation 
which is valid for the strong segregation region and for the macrophase
separation. In our previous work
\cite{Uneyama-Doi-2005} we proposed the free energy functional which
is valid for strong segregation region, that is, valid for
vesicles \cite{Uneyama-Doi-2005a}.

\subsection{Free Energy Functional}
\label{free_energy_functional}
The free energy functional for the system can be expressed as follows by using the
soft-colloid picture \cite{Louis-Bolhuis-Hansen-Meijer-2000,Pagonabarraga-Cates-2001,Frusawa-2005}.
\begin{equation}
 \label{freeenergy_softcolloid}
  F = - T S + U
  = - T S_{\text{trans}} - T S_{\text{conf}} + U
\end{equation}
where $S$ and $U$ are the entropy functional and the interaction energy
functional, and $T$ is the temperature. The entropy functional can be
decomposed into two parts; the
translational entropy functional $S_{\text{trans}}$ and the
conformational entropy functional $S_{\text{conf}}$.
In the density functional theory, $U$, $S_{\text{trans}}$ and
$S_{\text{conf}}$ are expressed as the functional of the local volume
fraction fields. (The local volume fraction field is equivalent to 
the local density field, if the segment volume is
set to unity. We assume that the segment volume is unity in this work.)

Each contributions to the free energy functional can be modelled for AB
type diblock copolymer and solvent mixtures by
\cite{Uneyama-Doi-2005,Uneyama-Doi-2005a}
\begin{equation}
 \label{translational_entropy}
  - \frac{S_{\text{trans}}}{k_{B}} =
  \sum_{i \, (= A,B)} \int d\bm{r} \, f_{i} C_{ii} \phi_{i}(\bm{r}) \ln \phi_{i}(\bm{r})
  + \int d\bm{r} \, \phi_{S}(\bm{r}) \ln \phi_{S}(\bm{r}) 
\end{equation}
\begin{equation}
 \label{conformational_entropy}
 \begin{split}
  - \frac{S_{\text{conf}}}{k_{B}} = &
  \sum_{i,j \, (= A,B)}  \int d\bm{r} d\bm{r}' \, 2 \sqrt{f_{i} f_{j}} A_{ij}
  \tilde{\mathcal{G}}(\bm{r} - \bm{r}') \psi_{i}(\bm{r}) \psi_{j}(\bm{r}') \\
  & + \int d\bm{r} \, 4 \sqrt{f_{A} f_{B}} C_{AB} \psi_{A}(\bm{r}) \psi_{B}(\bm{r})
  + \sum_{i \, (= A,B,S)} \int d\bm{r} \, \frac{b^{2}}{6} \left| \nabla \psi_{i}(\bm{r}) \right|^{2}
 \end{split}
\end{equation}
\begin{equation}
 \label{interaction_energy}
  \frac{U}{k_{B} T} = \sum_{i,j \, (= A,B,S)} \int d\bm{r} \, \frac{\chi_{ij}}{2} \phi_{i}(\bm{r}) \phi_{j}(\bm{r})
  + \int d\bm{r} \frac{P(\bm{r})}{2} \left[ \phi_{A}(\bm{r}) + \phi_{B}(\bm{r}) + \phi_{S}(\bm{r}) - 1 \right]
\end{equation}
where $k_{B}$ is the Boltzmann constant,
$\phi_{i}(\bm{r})$ is the local volume fraction of the $i$ segment at
position $\bm{r}$, $\psi_{i}(\bm{r})$ is the order parameter field ($\psi$-field)
defined as $\psi_{i}(\bm{r}) \equiv \sqrt{\phi_{i}(\bm{r})}$ \cite{Frusawa-2005}.
The coefficients $A_{ij}$ and $C_{ij}$ are constants determined from the block
copolymer architecture (we don't show the explicit forms of $A_{ij}$ and
$C_{ij}$ here; they can be found in Refs
\citen{Uneyama-Doi-2005a,Uneyama-Doi-2005}), $f_{i}$ is the block
ratio of the block copolymer, $b$ is the Kuhn length, $\chi_{ij}$ is
the Flory-Huggins $\chi$ parameter, and
$\tilde{\mathcal{G}}(\bm{r} - \bm{r}')$ is the Green function
which satisfies $[ -\nabla^{2} + \lambda^{-2} ] \tilde{\mathcal{G}}(\bm{r}
- \bm{r}') = \delta(\bm{r} - \bm{r}')$
\cite{Tang-Freed-1992,notes-inpreparation},
where $\lambda$ is a cutoff length and is about the size of
microphase separation structures.
$P(\bm{r})$ is the Lagrange multiplier for the
incompressible condition ($\phi_{A}(\bm{r}) + \phi_{B}(\bm{r}) +
\phi_{S}(\bm{r}) = 1$) \cite{Drolet-Fredrickson-1999}.

The conformational entropy \eqref{conformational_entropy} is expressed
in the bilinear form of $\psi$-field
\cite{Lifshitz-Grosberg-Khokhlov-1978,Grosberg-Khokhlov-book,Uneyama-Doi-2005}.
One significant property of the conformational entropy
is that it satisfies the following relation.
\begin{equation}
 \label{conformational_entropy_extensivity}
 S_{\text{conf}}\left[ \{ \alpha \phi_{i}(\bm{r}) \} \right]
  = \alpha S_{\text{conf}}\left[ \{ \phi_{i}(\bm{r}) \} \right]
\end{equation}
where $\alpha$ is arbitrary positive constant.
(It is clear that the conformational entropy in the bilinear form of $\psi$
satisfies eq \eqref{conformational_entropy_extensivity}.)
We can interpret eq \eqref{conformational_entropy_extensivity} as
the extensivity of the conformational entropy; under the current
approximation (using mean field, and neglecting many body correlations),
the total conformational entropy is sum of the
conformational entropy of each polymer chains. Since the density profiles
of miecellar structures are determined to achieve the low free energy,
the conformational entropy which does not satisfy eq
\eqref{conformational_entropy_extensivity} may lead unphysical density
profile. We note that the Lifshitz entropy
\cite{Lifshitz-Grosberg-Khokhlov-1978,Grosberg-Khokhlov-book} and the
conformational entropy calculated by the SCF theory satisfy eq \eqref{conformational_entropy_extensivity},
while most of previous phenomenological
free energy models do not satisfy eq \eqref{conformational_entropy_extensivity}.

Substituting eqs \eqref{translational_entropy} -
\eqref{interaction_energy} into eq \eqref{freeenergy_softcolloid}, we get
\begin{equation}
 \label{freeenergy_diblocksolution}
 \begin{split}
  \frac{F}{k_{B} T} = 
  & \sum_{i \, (= A,B)} \int d\bm{r} \, 2 f_{i} C_{ii} \psi^{2}_{i}(\bm{r}) \ln \psi_{i}(\bm{r})
    + \int d\bm{r} \, 2 \psi^{2}_{S}(\bm{r}) \ln \psi_{S}(\bm{r}) \\
  & + \sum_{i,j \, (= A,B)}  \int d\bm{r} d\bm{r}' \, 2 \sqrt{f_{i} f_{j}} A_{ij}
  \tilde{\mathcal{G}}(\bm{r} - \bm{r}') \psi_{i}(\bm{r}) \psi_{j}(\bm{r}') \\
  & + \int d\bm{r} \, 4 \sqrt{f_{A} f_{B}} C_{AB} 
  \psi_{A}(\bm{r}) \psi_{B}(\bm{r}) 
  + \sum_{i \, (= A,B,S)} \int d\bm{r} \, \frac{b^{2}}{6} \left| \nabla \psi_{i}(\bm{r}) \right|^{2} \\
  & + \sum_{i,j \, (= A,B,S)} \int d\bm{r} \, \frac{\chi_{ij}}{2} \psi^{2}_{i}(\bm{r}) \psi^{2}_{j}(\bm{r})
  + \int d\bm{r} \frac{P(\bm{r})}{2} \left[ \psi^{2}_{A}(\bm{r}) + \psi^{2}_{B}(\bm{r}) + \psi^{2}_{S}(\bm{r}) - 1 \right]
 \end{split}
\end{equation}
The difference between the free energy functional
\eqref{freeenergy_diblocksolution} and our previous model
\cite{Uneyama-Doi-2005} is the form of the Green function
(in the previous theory, the Green function $\mathcal{G}(\bm{r} -
\bm{r}')$ has no cutoff, $- \nabla^{2} \mathcal{G}(\bm{r} - \bm{r}') =
\delta(\bm{r} - \bm{r}')$).
A cutoff for the Green function was first introduced
by Tang and Freed \cite{Tang-Freed-1992}, without derivation, for the Ohta-Kawasaki type
standard GL free energy \cite{Ohta-Kawasaki-1986}.
The Ohta-Kawasaki type Coulomb type long range interaction term may
cause unphysical
interaction or correlation in non-periodic systems such as block
copolymer solutions. The true form of long range interaction can be, in
principle, obtained by calculating higher order terms in free energy
functional (the effect of
higher order vertex functions) exactly.
While we can calculate the higher order terms numerically by using the
SCF, it is practically impossible to get analytical form of them and
take a summation.

To avoid the unphysical correlation without numerically demanding
calculations, in the present study,
We employ a theory which describes more precisely
the properties of the length scale of the order of one polymer chain
\cite{notes-inpreparation};
We modify the Coulomb type interaction by introducing Tang-Freed type
cutoff \cite{Tang-Freed-1992}, instead of calculating higher order
terms exactly.
It is reasonable to consider that
the interaction range (or the cutoff length $\lambda$) of the long range
interaction cannot exceed
the characteristic size of the polymer chain.
Generally it is difficult to determine the
cutoff length (one way to determine it is to use the SCF calculation or
other microscopic calculations such as the MD).

Here we estimate the cutoff length for the simplest case, diblock
copolymer melts.
For homogeneous ideal state, the cutoff length is considered to be
the mean square end to end distance of a polymer chain $N^{1/2} b$ where
$N$ is the polymerization index.
On the other hand, for microphase separation structures at the strong
segregation region, the polymer chains are strongly stretched.
The periods of the structures are known to be
proportional to $N^{2/3} b$
\cite{Ohta-Kawasaki-1986,Hashimoto-Shibayama-Kawai-1980}.
Thus we have
the following polymerization index dependence for the cutoff length of diblock copolymer melts.
\begin{equation}
 \label{lambda_melt}
 \lambda \propto
  \begin{cases}
   N^{1/2} b & (\text{for ideal state}) \\
   N^{2/3} b & (\text{for strong segregation region})
  \end{cases}
\end{equation}

We assume that the micellar structures have the similar cutoff length as
the melt case.
In this work we use $\lambda \simeq N^{2/3} b$ for micellar systems.
Note that this is a rough estimation and the validity
is not guaranteed.
For example, the proportional coefficient in eq \eqref{lambda_melt} is
ignored here, or the effect of swelling of the hydrophilic subchain in the
solution is not taken into account.
Therefore the validity of this value of $\lambda$ should be tested by
comparing with the characteristic scale of the resulting phase
separation structure. We also note that the
dependence of the morphologies to the value of $\lambda$ is not so
sensitive (see also Appendix \ref{cutoff_length_dependency}).
What important here is that the interaction range is not
infinite (as the original Ohta-Kawasaki green function) but finite
(as the modified green function by Tang and Freed).

\subsection{Dynamic Equation}
\label{dynamic_equation}
We employ the stochastic dynamic density functional model
\cite{Dean-1996,Frusawa-Hayakawa-2000,Archer-Rauscher-2004} for the time
evolution equation.
The diblock copolymer solutions are expressed as three component systems
(hydrophilic subchain $A$, hydrophobic subchain $B$, and solvent $S$).
The dynamic equation for multicomponent systems can be expressed as follows.
\begin{equation}
 \label{dynamic_density_functional_equation}
  \frac{\partial \phi_{i}(\bm{r})}{\partial t} = 
  \nabla \cdot \left[ \frac{1}{\zeta_{i}} \phi_{i}(\bm{r}) \nabla \frac{\delta F}{\delta \phi_{i}(\bm{r})}\right] + \xi_{i}(\bm{r},t)
\end{equation}
where $\phi_{i}(\bm{r})$ is the concentration of $i$-th component ($i = A,B,S$)
and $\zeta_{i}$ is a friction coefficient.
$F$ is the free energy. $\xi_{i}(\bm{r},t)$ is the
thermal noise which satisfies the
fluctuation-dissipation relation.
\begin{align}
 \label{fluctuation_dissipation_relation_first}
 \left\langle \xi_{i}(\bm{r},t) \right\rangle & = 0 \\
 \label{fluctuation_dissipation_relation_second}
 \left\langle \xi_{i}(\bm{r},t) \xi_{j}(\bm{r}',t') \right\rangle & =
 - \frac{2}{\zeta_{i}} \tilde{\beta}^{-1} k_{B} T \delta_{ij} \nabla \cdot
 \left[ \phi_{i}(\bm{r}) \nabla \delta(\bm{r} - \bm{r}') \right]
 \delta(t - t')
\end{align}
where $\langle \dots \rangle$ represents the statistical average and 
$0 \le \tilde{\beta}^{-1} < 1$ is the constant determined
from the degree of coarse graining (see
Ref \citen{Archer-Rauscher-2004} for detail). One can interpret that the
temperature of the noise is the effective temperature
$\tilde{\beta}^{-1} T$, instead of the real temperature $T$.
We also note that $\xi_{i}(\bm{r})$ can be generated easily by the algorithm
proposed by van Vlimmeren and Fraaije \cite{vanVlimmeren-Fraaije-1996}.
It should be noted here that the free energy functional in eq
\eqref{dynamic_density_functional_equation} is generally some kind of
effective free energy functional, and is not equal to
the free energy functional for the equilibrium state
\cite{Archer-Rauscher-2004}. In this work we approximate the effective
free energy functional as the equilibrium free energy functional
\eqref{freeenergy_diblocksolution}. This approximation corresponds to
assume that all the polymer chains are fully relaxed, and thus this
approximation neglects the viscoelasticity associated with the
relaxation of polymer chains.

We can interpret the dynamic equation \eqref{dynamic_density_functional_equation} as
the TDGL equation. In this case, the Onsager coefficients (or the
mobility) $\phi_{i}(\bm{r}) / \zeta_{i}$ 
is proportional to $\phi_{i}(\bm{r})$. This is most
important for the strong segregation region or in the situation that the
solute concentration is sufficiently small
\cite{Langer-Baron-Millar-1975,Kitahara-Imada-1978,deGennes-1980}.
Here we note that the TDGL equation with density dependent mobility
and the free energy of ideal gases (the translational entropy
of ideal gases) reduces to the diffusion equation (or the Smoluchowski equation)
\cite{Doi-Edwards-book}.
This suggests that we should use the Flory-Huggins (or Bragg-Williams)
type free energy model once we employed the density dependent mobility.
Similarly we should use the density dependent mobility 
if we employ the Flory-Huggins type free energy model. It is also noted
that the TDGL equation with the constant mobility and the Flory-Huggins
type free energy leads unphysical singular behaviour near
$\phi_{i}(\bm{r}) = 0$.

For simplicity, we assume that all the segments have the same friction
coefficient ($\zeta_{i} = \zeta$).
We can rewrite eq \eqref{dynamic_density_functional_equation}
by using $\phi_{i}(\bm{r}) = \psi^{2}_{i}(\bm{r})$ as follows.
\begin{equation}
 \label{dynamic_density_functional_equation_psi}
  \begin{split}
   \frac{\partial \phi_{i}(\bm{r})}{\partial t} 
   & = \nabla \cdot \left[ \frac{k_{B} T}{\zeta} \psi^{2}_{i}(\bm{r}) \nabla \left[ \frac{\delta (F / k_{B} T)}{\delta \psi_{i}(\bm{r})} \frac{\partial \psi_{i}(\bm{r})}{\partial \psi^{2}_{i}(\bm{r})} \right] \right] + \xi_{i}(\bm{r},t) \\
   & = \frac{k_{B} T}{2 \zeta} \left[ \psi_{i}(\bm{r}) \nabla^{2} \mu_{i}(\bm{r}) - \mu_{i}(\bm{r}) \nabla^{2} \psi_{i}(\bm{r}) \right] + \xi_{i}(\bm{r},t) \\
   & = \frac{k_{B} T}{2 \zeta} \left[ \psi_{i}(\bm{r}) \nabla^{2} \mu_{i}(\bm{r}) - \mu_{i}(\bm{r}) \nabla^{2} \psi_{i}(\bm{r}) + \frac{2 \zeta}{k_{B} T} \xi_{i}(\bm{r},t) \right]
  \end{split}
\end{equation}
where $\mu_{i}(\bm{r}) \equiv \delta (F / k_{B} T) / \delta \psi_{i}(\bm{r})$ is a
kind of chemical potential field.
It is noted that $\mu_{i}(\bm{r})$ is not
singular at $\psi_{i}(\bm{r}) = 0$ and thus we can perform stable simulations
(on the other hand, $\delta (F / k_{B} T) / \delta \phi_{i}(\bm{r})$ has a singularity
and is numerically unstable).

By introducing rescaled variables defined as
\begin{align}
 \label{tilde_t_definition}
 \tilde{t} & \equiv \frac{k_{B} T}{2 \zeta} t \\
 \label{tilde_xi_definition}
 \tilde{\xi}_{i}(\bm{r},t) & \equiv \frac{2 \zeta}{k_{B} T} \xi_{i}(\bm{r}, t)
\end{align}
and substituting eqs
\eqref{tilde_t_definition} and \eqref{tilde_xi_definition} into
eqs \eqref{dynamic_density_functional_equation_psi},
\eqref{fluctuation_dissipation_relation_first} and \eqref{fluctuation_dissipation_relation_second},
we obtain the dynamic equation for rescaled variables
\begin{equation}
 \label{dynamic_density_functional_equation_psi_rescaled}
   \frac{\partial \phi_{i}(\bm{r})}{\partial \tilde{t}}
   = \psi_{i}(\bm{r}) \nabla^{2} \mu_{i}(\bm{r}) - \mu_{i}(\bm{r}) \nabla^{2} \psi_{i}(\bm{r}) + \tilde{\xi}_{i}(\bm{r},\tilde{t})
\end{equation}
and the fluctuation dissipation relation for the rescaled noise field
\begin{align}
 \label{fluctuation_dissipation_relation_rescaled_first}
 \left\langle \tilde{\xi}_{i}(\bm{r},\tilde{t}) \right\rangle & = 0 \\
 \label{fluctuation_dissipation_relation_rescaled_second}
 \left\langle \tilde{\xi}_{i}(\bm{r},\tilde{t}) \tilde{\xi}_{j}(\bm{r}',\tilde{t}') \right\rangle & =
 - 4 \tilde{\beta}^{-1} \delta_{ij} \nabla \cdot
 \left[ \phi_{i}(\bm{r}) \nabla \delta(\bm{r} - \bm{r}') \right]
 \delta(\tilde{t} - \tilde{t}')
\end{align}
It should be noted here that the magnitude of the noise $\tilde{\xi}_{i}(\bm{r},\tilde{t})$
depends only on $\tilde{\beta}^{-1}$. The temperature change corresponds
to the change of the time scale (since $t \propto \tilde{t} / T$)
and the change of the $\chi$ parameter.
However, notice that to return from the to rescaled time $\tilde{t}$ (which is used
in the actual simulations) to the real
time scale $t$, we have to multiply the factor 
$2 \zeta / k_{B} T$.

We note that we can set $k_{B} T / 2 \zeta = 1$ instead of
introducing rescaled variables ($\tilde{t}$ and $\tilde{\xi}_{i}$).
This corresponds to setting the effective diffusion coefficient for the monomer
to unity (strictly speaking, the half of the effective diffusion
coefficient is set to unity).
This change can be done without losing generality, as shown above.
To return to the real time scale, we have to multiply the factor $2
\zeta / k_{B} T$ to the rescaled time. This factor can be estimated from
the experimental diffusion data. (We will estimate the real time scale
in Section \ref{comparison_with_experiments}.)

It should be noted here that
eqs \eqref{dynamic_density_functional_equation_psi_rescaled} and
 \eqref{fluctuation_dissipation_relation_rescaled_second}
implies that the annealing process (the temperature change process)
corresponds to the change of the
$\chi$ parameter, and the magnitude of the noise is not changed because
there are no other parameters related to $T$. The magnitude of the noise
is characterized only by $\tilde{\beta}^{-1}$ in the rescaled units.

%------------------------------------------------------------------------------
\section{Simulation}
\label{simulation}
We solve eq
\eqref{dynamic_density_functional_equation_psi_rescaled} numerically in three dimensions.
 The chemical potential
$\mu_{i}(\bm{r})$ can be calculated from the free energy functional
\eqref{freeenergy_diblocksolution} whereas thermal noise $\xi_{i}(\bm{r},t)$
is calculated by the van Vlimmeren and Fraaije's algorithm
\cite{vanVlimmeren-Fraaije-1996}.

\subsection{Numerical Scheme}
\label{numerical_scheme}
Simulations are started from the homogeneous state
($\phi_{i}(\bm{r}) = \bar{\phi}_{i}$, where $\bar{\phi}_{i}$ is the
spatial average of $\phi_{i}(\bm{r})$).
Each step of the time evolution in the simulation is
as follows;
\begin{enumerate}
 \item \label{simulation_firststep}
       Calculate the $\psi$-field $\psi_{i}(\bm{r})$ and the chemical
       potential field $\mu_{i}(\bm{r})$ from the density field $\phi_{i}(\bm{r})$.
 \item Generate the thermal noise $\tilde{\xi}_{i}(\bm{r},\tilde{t})$.
 \item Calculate the Lagrange multiplier $P(\bm{r})$.
 \item \label{density_time_evolution}
       Evolve the density field $\phi_{i}(\bm{r})$ by time step
       ${\Delta}{\tilde t}$, using $\psi_{i}(\bm{r}), \mu_{i}(\bm{r})$ and $\tilde{\xi}_{i}(\bm{r},\tilde{t})$.
 \item Return to the step \ref{simulation_firststep}.
\end{enumerate}

Here we describe the numerical scheme for the simulations in detail.
As mentioned above, the dynamic density functional equation
\eqref{dynamic_density_functional_equation} is reduced to the diffusion
equation for the Flory-Huggins type translational entropy functional.
In our free energy \eqref{freeenergy_diblocksolution}, we have the
Flory-Huggins translational entropy term. Thus the dynamic equation
\eqref{dynamic_density_functional_equation_psi_rescaled} contains the
standard diffusion term (Laplacian term).
\begin{equation}
 \frac{\partial \phi_{i}(\bm{r})}{\partial \tilde{t}} = 2 \tilde{C}_{i} \nabla^{2} \phi_{i}(\bm{r}) + \dotsb
\end{equation}
where we set $\tilde{C}_{i} = f_{i} C_{ii}$ ($i = A,B$) and $\tilde{C}_{s} = 1$.
The numerical stability of the diffusion type equation can be improved
by the implicit schemes. In the previous work \cite{Uneyama-Doi-2005a}
we employed the alternating direction implicit (ADI) scheme 
\cite{NumericalRecipes} for the Laplacian term. In this work we 
employ the ADI scheme to improve stability, as the numerical scheme for
the equilibrium simulations.

We split the each time evolution steps (time step $\Delta\tilde{t}$) into
three substeps (time step $\Delta\tilde{t} / 3$).
The update scheme for the density field from $\phi^{(n)}_{i}(\bm{r})
\equiv \phi_{i}(\bm{r},\tilde{t} + n \Delta\tilde{t} / 3)$ to
$\phi^{(n + 1)}_{i}(\bm{r})$ is as follows.
(For simplicity, here we use the continuum expression for the position
$\bm{r}$ and the differential operators
$\partial^{2} / \partial x^{2}, \partial^{2} / \partial y^{2},
\partial^{2} / \partial z^{2}$ and $\nabla^{2}$. In real simulations we
use the standard lattice for the position and the center difference
operators for the differential operators \cite{NumericalRecipes})
\begin{align}
 \label{update_scheme_adi_x}
 \begin{split}
 \left[ 1 - \frac{2 \Delta\tilde{t}}{3} \tilde{C}_{i} \frac{\partial^{2}}{\partial x^{2}} \right] \phi^{(1)}_{i}(\bm{r})
 = & \, \phi^{(0)}_{i}(\bm{r}) + \frac{\Delta\tilde{t}}{3} \bigg[ - 2 \tilde{C}_{i} \frac{\partial^{2}}{\partial x^{2}} \phi^{(0)}_{i}(\bm{r}) + \psi^{(0)}_{i}(\bm{r}) \nabla^{2} \mu^{(0)}_{i}(\bm{r}) \\
 & \qquad \qquad \qquad - \mu^{(0)}_{i}(\bm{r}) \nabla^{2} \psi^{(0)}_{i}(\bm{r}) + \tilde{\xi}^{(0)}_{i}(\bm{r},\tilde{t}) \bigg]
 \end{split} \\
 \label{update_scheme_adi_y}
 \begin{split}
 \left[ 1 - \frac{2 \Delta\tilde{t}}{3} \tilde{C}_{i} \frac{\partial^{2}}{\partial y^{2}} \right] \phi^{(2)}_{i}(\bm{r})
 = \, & \phi^{(1)}_{i}(\bm{r}) + \frac{\Delta\tilde{t}}{3} \bigg[ - 2 \tilde{C}_{i} \frac{\partial^{2}}{\partial y^{2}} \phi^{(1)}_{i}(\bm{r}) + \psi^{(1)}_{i}(\bm{r}) \nabla^{2} \mu^{(1)}_{i}(\bm{r}) \\
 & \qquad \qquad \qquad - \mu^{(1)}_{i}(\bm{r}) \nabla^{2} \psi^{(1)}_{i}(\bm{r}) + \tilde{\xi}^{(1)}_{i}(\bm{r},\tilde{t}) \bigg]
 \end{split} \\
 \label{update_scheme_adi_z}
 \begin{split}
 \left[ 1 - \frac{2 \Delta\tilde{t}}{3} \tilde{C}_{i} \frac{\partial^{2}}{\partial z^{2}} \right] \phi^{(3)}_{i}(\bm{r})
 = \, & \phi^{(2)}_{i}(\bm{r}) + \frac{\Delta\tilde{t}}{3} \bigg[ - 2 \tilde{C}_{i} \frac{\partial^{2}}{\partial z^{2}} \phi^{(2)}_{i}(\bm{r}) + \psi^{(2)}_{i}(\bm{r}) \nabla^{2} \mu^{(2)}_{i}(\bm{r}) \\
 & \qquad \qquad \qquad - \mu^{(2)}_{i}(\bm{r}) \nabla^{2} \psi^{(2)}_{i}(\bm{r}) + \tilde{\xi}^{(2)}_{i}(\bm{r},\tilde{t}) \bigg]
 \end{split}
\end{align}
In each steps, the chemical potential $\mu_{i}^{(n)}(\bm{r})$ and
$\xi_{i}^{(n)}(\bm{r})$ is regenerated.
The numerical Scheme for the calculation of the chemical potential
$\mu_{i}^{(n)}(\bm{r})$ is found in Ref \citen{Uneyama-Doi-2005a}.
Eqs \eqref{update_scheme_adi_x} - \eqref{update_scheme_adi_z} can be
solved easily by using the numerical scheme for the cyclic tridiagonal
matrix \cite{NumericalRecipes}.

We need to calculate $\psi^{(n)}_{i}(\bm{r})$ to calculate the chemical
potential $\mu_{i}^{(n)}(\bm{r})$. Because of the thermal noise and numerical error, the condition
$0 \le \phi^{(n)}_{i}(\bm{r}) \le 1$ is not always satisfied.
Thus we put $\psi^{(n)}_{i}(\bm{r})$ as
\begin{equation}
 \psi^{(n)}_{i}(\bm{r}) =
  \begin{cases}
   0 & (\phi^{(n)}_{i}(\bm{r}) < 0) \\
   1 & (\phi^{(n)}_{i}(\bm{r}) > 1) \\
   \sqrt{\phi^{(n)}_{i}(\bm{r})} & (\text{otherwise})
  \end{cases}
\end{equation}
This avoids numerical difficulty associated with negative value of
$\phi^{(n)}_{i}(\bm{r})$.

The noise field $\xi^{(n)}_{i}(\bm{r},\tilde{t})$ is calculated at each
ADI time steps, by using the scheme described in Appendix
\ref{noise_generation_scheme} (notice that the size of the time step
here is $\Delta\tilde{t} / 3$ instead of $\Delta\tilde{t}$).

The Lagrange
multiplier $P(\bm{r})$ is also updated at each ADI time steps by the following scheme.
Because of the large thermal noise, the incompressible condition is
not always satisfied. Thus we use roughly approximated and relatively simple
scheme to calculate $P(\bm{r})$.
If we ignore the terms in the chemical potential except for the Lagrange
multiplier term, the ADI update scheme can be approximately written as
\begin{equation}
 \label{update_scheme_p}
  \phi^{(n + 1)}_{i}(\bm{r})
  \approx \phi^{(n)}_{i}(\bm{r}) + \frac{\Delta\tilde{t}}{3} \nabla \cdot \left[ \phi^{(n)}_{i}(\bm{r}) \nabla P(\bm{r}) \right]
\end{equation}
From eq \eqref{update_scheme_p} we have
\begin{equation}
 \begin{split}
  \sum_{i} \phi^{(n + 1)}_{i}(\bm{r})
  & \approx \sum_{i} \phi^{(n)}_{i}(\bm{r}) + \frac{\Delta\tilde{t}}{3} \nabla \cdot \left[ \sum_{i} \phi^{(n)}_{i}(\bm{r}) \nabla P(\bm{r}) \right] \\
  \frac{1}{\Delta\tilde{t} / 3} \left[ \sum_{i} \phi^{(n + 1)}_{i}(\bm{r}) - \sum_{i} \phi^{(n)}_{i}(\bm{r}) \right]
  & \approx \nabla^{2} P(\bm{r}) - \nabla \cdot \left[ \left[ 1 - \sum_{i} \phi^{(n)}_{i}(\bm{r}) \right] \nabla P(\bm{r}) \right]
 \end{split}
\end{equation}
Using the constraint $\sum_{i} \phi^{(n + 1)}_{i}(\bm{r}) = 1$ and
assuming $|1 - \sum_{i} \phi^{(n)}(\bm{r})| \ll 1$,
finally we have
\begin{equation}
 \label{update_scheme_kappa}
 \nabla^{2} P(\bm{r}) \approx \frac{1}{(\Delta\tilde{t} / 3)}  \left[ 1 - \sum_{i} \phi^{(n)}_{i}(\bm{r}) \right]
\end{equation}
Eq \eqref{update_scheme_kappa} and the long range interaction terms
(which contains the Green function $\tilde{\mathcal{G}}(\bm{r} - \bm{r}')$) in
the chemical potential $\mu_{i}(\bm{r})$ can be calculated by several
numerical methods for partial differential equations. Here we calculate them by
the direct method, using FFTW3 \cite{Frigo-Johnson-2005}.

\subsection{Results}
\label{results}
Parameters for the diblock copolymer are as follows: polymerization index
$N = 10$, block ratio $f_{A} = 1/3, f_{B} = 2/3$, the spatial average of the
volume fraction
$\bar{\phi}_{p} = 0.2$.
The cutoff length for
the long range interaction should be set appropriately. 
If there are
no solvent diblock copolymers form strongly segregated
microphase separation structure (currently we are interested in the strong
segregation region).
Thus from eq \eqref{lambda_melt} we
have $\lambda \propto N^{2/3} b \simeq 4.64$.
Here we set $\lambda = 5$.
The validity of this value is argued later.
(See also Appendix \ref{cutoff_length_dependency}.)

Parameters for the solvent are as follows:
the spatial average of the volume fraction $\bar{\phi}_{s} = 1 - \bar{\phi}_{p} = 0.8$.
Flory-Huggins $\chi$ parameters are $\chi_{AB} = 2.5, \chi_{AS} = -0.5, \chi_{BS} = 5$
(the $A$ monomer is hydrophilic and the $B$ monomer is hydrophobic) and
the Kuhn length is set to unity ($b = 1$).
All the simulations are carried out in three dimensions. The size of the
simulation box is $24b \times 24b \times 24b$ and the number of lattice
points is $48 \times 48 \times 48$. We apply the periodic boundary condition.
The time step is set to ${\Delta}{\tilde t} = 0.0025$ and the magnitude of
noise is
$\tilde{\beta}^{-1} =  0.0390625$.
The simulation is carried out up to $2500000$ time
steps (from $\tilde{t} = 0$ to $\tilde{t} = 6250$) and it
requires about $20$ days on a $2.8\text{GHz}$
Xeon work station.

The snapshots of dynamics simulations are shown in
Figure \ref{vesicle_formation_snapshots}.
The observed vesicle formation process is different from the mechanism II,
which is observed in the previous thermal equilibrium
simulations \cite{He-Liang-Huang-Pan-2004,Uneyama-Doi-2005a}
and EPD simulations \cite{He-Schmid-2006}.
First, spherical micelles are formed (Figure
\ref{vesicle_formation_snapshots}(a)). Then the micelles aggregate 
and become large (Figure
\ref{vesicle_formation_snapshots}(b)), and a disk-like micelle (Figure
\ref{vesicle_formation_snapshots}(c)) appears. The disk-like micelle is closed (Figure
\ref{vesicle_formation_snapshots}(d),(e)) to form a vesicle
(Figure \ref{vesicle_formation_snapshots}(f)).
This vesicle formation
process agrees with the mechanism I, which is observed in the particle simulations
\cite{Noguchi-Takasu-2001,Yamamoto-Maruyama-Hyodo-2002,Marrink-Mark-2003}.

To confirm that our vesicle formation is independent of the random seed,
next we perform simulations with different random
seeds. We peform four simulations here and vesicles are obtained in two
simulations.
Figures \ref{vesicle_formation_snapshots2} and
\ref{vesicle_formation_snapshots3} show the snapshots of vesicle
formation processes obtained for different random seeds (all other
parameters are the same as the previous simulation).
The vesicle formation processes in Figures
\ref{vesicle_formation_snapshots2} and
\ref{vesicle_formation_snapshots3} again agree with the mechanism I.
While the vesicles are not observed for all the simulations,
but we consider that the vesicle formation process obtained by
the previous simulation is not artifact.

The characteristic size of the phase separation structure
should be compared with the cutoff length $\lambda$.
The characteristic size of the micellar structure can be found in the
two dimensional cross
section data or the density correlation function.
Figure \ref{cross_section_density_correlation}(a) and (b) show the
two dimensional cross section of the final structure ($\tilde{t} = 6250$, Figure
\ref{vesicle_formation_snapshots}(f)).
From Figure \ref{cross_section_density_correlation}(a) and (b) we can
observe that the cutoff length $\lambda = 5$ is comparable to the
characteristic size of the micelle bilayer structure.
Figure \ref{cross_section_density_correlation} (c) shows the radial averaged
correlation function of the density field in the Fourier space $S_{AB}(q)$,
defined as \cite{Shinozaki-Oono-1993}
\begin{equation}
 S_{AB}(q) \equiv \frac{\displaystyle \sum_{q - \Delta / 2 \le |\bm{q}'| < q + \Delta / 2} \left[ \phi_{A}(\bm{q'}) / f_{A} - \phi_{B}(\bm{q'}) / f_{B} \right]^{2}}
  {\displaystyle \sum_{q - \Delta / 2 \le |\bm{q}'| < q + \Delta / 2} 1}
\end{equation}
where $\phi_{i}(\bm{q})$ is the Fourier transform of the density field
and $\Delta$ is the width of the shell in the Fourier space (here we set
$\Delta = 0.5 \times 2 \pi$).
The correlation functions
have peaks at $q / 2 \pi \simeq 0.15$. Thus the characteristic wave length
for the corona (the $A$ subchain rich region) and the core (the $B$
subchain rich region) is $\simeq 6.7$. While the resolution (the shell
width $\Delta$) is not fine, the characteristic size of the structure is
comparable with the value of the cutoff length, $\lambda = 5$.
Therefore we consider that the value of $\lambda$ used in this
simulation is appropriate and valid. (See also Appendix \ref{cutoff_length_dependency}).

Figure \ref{structure_volume_fraction_dependence} shows various
micellar structures obtained from simulations for different volume fractions
$\bar{\phi}_{p} =
0.1, 0.15, 0.25,$ and $0.3$.
Simulations are performed up to $1250000$ time steps (from $\tilde{t} = 0$ to $\tilde{t} = 6250$).
All other parameters are the same as those
in Figure \ref{vesicle_formation_snapshots}.
If the volume fraction of the polymer is small (Figure
\ref{structure_volume_fraction_dependence}(a),(b)) only small micelles
(spherical micelles in \ref{structure_volume_fraction_dependence}(a) and
disk-like micelles in \ref{structure_volume_fraction_dependence}(b)) are
formed. On the other hand, if the volume fraction of the polymer is large
(Figure \ref{structure_volume_fraction_dependence}(c),(d)) bilayer
structures are formed. Vesicles in Figure
\ref{vesicle_formation_snapshots} are observed only for the intermediate
volume fraction.

In experiments, the mixed solvent (mixture of organic solvents and water) is often used
to control the morphology of the micellar structures of block copolymers
\cite{Shen-Eisenberg-1999,Shen-Eisenberg-2000,Bhargava-Zheng-Li-Quirk-Harris-Cheng-2006}.
If we assume that the mixed solvent is the mixture of common solvent
($\chi_{AS}$ and $\chi_{BS}$ is sufficiently small or negative) and
water, the volume fraction of water mainly affect the interaction
between the hydrophobic subchain and the solvent. We change $\chi_{BS}$
to mimic these experiments.

Figure \ref{structure_chi_parameter_dependence} shows the results of
simulations for various $\chi_{BS}$ ($\chi_{BS} = 0.2, 0.25, 0.3$ and
$0.35$).
We can observe the spherical micelles are formed for low $\chi_{BS}$ and
cylindrical micelles are formed for high $\chi_{BS}$.
This agree with the experimental results qualitatively.
Shen and Eisenberg \cite{Shen-Eisenberg-1999,Shen-Eisenberg-2000}
reported that by controlling solvent condition, one can control the
morphology of the block copolymer micelles.
The control of the morphology is especially important for application
use such as the drug delivery system.
Here we mimic the solvent condition change by changing the $\chi$
parameter between the hydrophobic segment and the solvent, $\chi_{BS}$.
We change the $\chi$ parameter at $\tilde{t} = 6250$ (Figure
\ref{vesicle_formation_snapshots}(f)),
from $\chi_{BS} = 5$ to $\chi_{BS} = 2.5$ or $3$, and perform
the simulations up to $\tilde{t} = 9375$.
The snapshots at $\tilde{t} = 9375$ are shown in
Figure \ref{structural_transition}.
The vesicle is changed into spherical micelles (Figure
\ref{structural_transition}(a)) or cylindrical micelles
(Figure \ref{structural_transition}(b)). The resulting morphologies are
consistent with the morphologies obtained by the simulations started
from the homogeneous state (Figure
\ref{structure_chi_parameter_dependence}(b) and Figure \ref{structure_chi_parameter_dependence}(c)).

%------------------------------------------------------------------------------
\section{Discussion}
\label{discussion}
From the dynamics simulations based on the density functional theory for
amphiphilic diblock copolymers,
we could observe the time-evolution of spontaneous formation of vesicles.
We also observed the micellar structures depending on the volume fraction
of polymers.
The observed vesicle formation process is just the same as the results of
the previous particle model simulations (mechanism I). At the initial
stage of time evolution, rapid
formation of small spherical micelles is observed. After the small
micelles are formed, they aggregate each other by collision and become larger
micelles. Finally the large disk-like micelles are closed
and become vesicles. The late stage process, collision process and the
close-up process,
 is slower than the initial stage, as the mechanism I.

There are several important differences between our model and the
standard TDGL
model for the weak segregation limit. Here we mention some essential properties of our model.
First, our model can be applied for the strong segregation region, at
least qualitatively. This is especially important, since the vesicles are
observed in the strong segregation region.

Second, the interaction between the hydrophilic subchain and solvent
should be handled carefully. If the hydrophilic interaction
($\chi_{AS}$) is too small
or both subchains are hydrophobic,
the system undergoes macrophase separation and separates into block
copolymer rich region and solvent rich region. In such a situation,
vesicles are not formed and onion structures
\cite{Koizumi-Hasegawa-Hashimoto-1994,Ohta-Ito-1995,Uneyama-Doi-2005,Sevink-Zvelindovsky-2005}
are formed instead.

Third, we applied rather
large thermal noise to the system.
Recently Zhang and Wang \cite{Zhang-Wang-2006} have shown that the glass
transition temperature and the spinodal line (stability limit of the
disordered phase) for the microphase
separation in diblock copolymer melts are quite close. This
implies that microphase separated structures in the block copolymer
systems are intrinsically glassy at the strong segregation region.
Without sufficiently large and realistic thermal noise,
the system is completely trapped at intermediate metastable structures
(in most cases, spherical micelles) and thus thermodynamically stable
structures
(vesicles) will never be achieved. As mentioned above, the magnitude of
the noise is determined by the characteristic time scale and the factor
$\tilde{\beta}^{-1}$ is independent of the temperature. Thus the noise
is important for 
mesoscopic systems even if the temperature of the system is not high.

\subsection{Thermal Noise}
\label{effective_temparature_for_noise}
Here we discuss about the thermal noise in detail.
Intuitively, the large thermal noise is needed to overcome the free
energy barrier.
In the kinetic pathway of usual macrophase separation
processes (for example, phase separations in homopolymer blends),
there are not so large free energy barriers.
Therefore, in most cases, the simulations for such systems
work well with small thermal noise or without thermal noise.
On the other hand, in the kinetic pathway of vesicle / micelle formation
processes, there are various
large free energy barriers (for example, in the collision and coalescence
process of spherical micelles).
However, it is not clear and established how to determine the
magnitude of the thermal noise for such systems.

As mentioned, the thermal noise in eq
\eqref{dynamic_density_functional_equation}
is expressed by using the effective temperature $\tilde{\beta}^{-1} T$,
instead of the real temperature $T$ and $\tilde{\beta}^{-1}$ is
determined from the degree of coarse graining.
Intuitively this means that the coarse graining procedure decreases the
degree of the freedom of the field. The similar approach can be found in
the van Vlimmeren and Fraaije's theory
\cite{vanVlimmeren-Fraaije-1996}. They employed the noise scaling
parameter $\Omega$ to express the effective degree of freedom in the
simulation cell.
These two approaches are similar and in fact, we can 
write the relation between them.
\begin{equation}
 \label{omega_beta_inv_relation}
 \Omega = \frac{|\bm{h}_{x}| |\bm{h}_{y}| |\bm{h}_{z}|}{\tilde{\beta}^{-1}}
\end{equation}
where $\bm{h}_{\alpha}$ is the lattice vector.
From eq \eqref{omega_beta_inv_relation} we can calculate the noise
scaling parameter for our simulation. Using the parameters used in
Section \ref{results}, we have $\Omega = (0.5^{3}) / 0.0390625 = 3.2$.
It should be noticed that this value of the noise scaling
parameter is much smaller than the values used in most of previous works ($\Omega = 100$)
by Fraaije et al
\cite{Zvelindovsky-vanVlimmeren-Sevink-Maurits-Fraaije-1998,Zvelindovsky-vanVlimmeren-Sevink-Maurits-Fraaije-1998b,vanVlimmeren-Maurits-Zvelindovsky-Sevink-Fraaije-1999}.
However, since the thermal noise play an important role in the vesicle
formation process, we cannot underestimate $\tilde{\beta}^{-1}$.
One may consider that the mean field approximation and the free energy
functional \eqref{freeenergy_diblocksolution} does not hold for low
noise scaling parameter systems. It may be true, but as shown by
Dean\cite{Dean-1996}, the dynamic density functional equation holds 
even if there are not large number of particles. Thus here we believe
that the mean field approximation still holds for our systems.

To show the necessity of the large thermal noise in the simulations,
we perform several simulations with various values of $\tilde{\beta}^{-1}$.
We expect that the simulations for small boxes are sufficient to see the
effect of the magnitude of the thermal noise, since the effect is
large even for early stage of the dynamics. Thus we perform simulations for small
systems (system size $12b \times 12b \times 12b$, lattice points $24
\times 24 \times 24$) with different values of $\tilde{\beta}^{-1}$,
$\tilde{\beta^{-1}} = 0.00390625, 0.0390625, 0.390625$.
All the other parameters are set to the previous vesicle formation
simulations in Section \ref{results} (Figure \ref{vesicle_formation_snapshots}).
Figure \ref{thermal_noise_snapshots} shows the snapshots of the
simulations. For small thermal noise case (Figure
\ref{thermal_noise_snapshots}(a),(b)), we can observe many micelles with
sharp interfaces. For large thermal noise case used in Section 
\ref{results} (Figure \ref{thermal_noise_snapshots}(c),(d))
we can observe the collision-coalescence type coarsening process of
micelles, driven by the thermal noise. For larger thermal noise case
(Figure \ref{thermal_noise_snapshots}(e),(f)) no structures are
observed. (We consider this is because the thermal noise is too large.)
Thus we can conclude that the large (but not too large) thermal noise
is required. This result is consistent with the proposition of van Vlimmeren et al
\cite{vanVlimmeren-Postma-Huetz-Brisson-Fraaije-1996}.
They carried out two dimensional simulations
for dense amphiphilic triblock copolymer solutions, with the full
(large) noise and the reduced (small) noise and proposed that the real
systems should have some
intermediate noise (smaller than the full noise, but larger than the
reduced noise).

It should be noted that simulations with small thermal noise is
very difficult to perform for our systems, because they are
in the strong segregation region.
The simulation with the small noise shows strong lattice anisotropy.
This means that there are very sharp interfaces which
often leads numerical instability, and thus we should use finer lattices
for such systems.
On the other hand, we do not observe such strong lattice anisotropy in
the simulations with large noise.
We believe that the large thermal noise stirs the density fields and
stabilizes the simulation.

At the end of this section, we note that it is difficult to determine
the parameter $\tilde{\beta}^{-1}$ theoretically.
The estimation of the magnitude of the thermal noise is
the open problem.

\subsection{Comparison with Experiments}
\label{comparison_with_experiments}

Next we discuss about the relation between the simulation results and
the experiments. We used many parameters for the simulations and they
should be related to the real experimental parameters. Unfortunately,
because the accuracy of our theory is not high and many approximations
are involved, it is difficult to calculate the real parameters
from our simulation parameters.

The most important parameters are the Flory-Huggins $\chi$
parameters. The $\chi$ parameters are experimentally measured or
calculated from the solubility parameters.
The $\chi$ parameter between the hydrophilic subchain and the
hydrophobic subchain is in most cases sufficiently large to cause
the microphase separation. For example,
Bhargava et al \cite{Bhargava-Zheng-Li-Quirk-Harris-Cheng-2006}
calculated the $\chi$
parameter between polystyrene (PS), which can be used as the hydrophobic
monomer \cite{Choucair-Eisenberg-2003}, 
and water as $\chi_{\text{PS,water}} = 6.27$.
Dormidontova \cite{Dormidontova-2002} measured the $\chi$
parameter between poly(ethyrene oxide) (PEO) and water and reported
that the $\chi$ parameter can be expressed as $\chi_{\text{PEO,water}} =
-0.0615 + 70 / T$ ($T$ is the absolute temperature).
Lam and Goldbeck-Wood \cite{Lam-GoldbeckWood-2003} also measured
$\chi_{\text{PEO,water}}$ and obtained
$\chi_{\text{PEO,water}} = 1.35$. 
Xu et al \cite{Xu-Winnik-Riess-Chu-Croucher-1992} measured 
the interaction between PS and PEO and reported $\chi_{\text{PS,PEO}} =
0.02 \sim 0.03$.
Zhu et al \cite{Zhu-Cheng-Calhoun-Ge-Quirk-Thomas-Hsiao-Yeh-Lotz-2001} also measured $\chi_{\text{PS,PEO}}$ and
and obtained $\chi_{\text{PS,PEO}} =
-0.00705 + 21.3 / T$. The polymerization index of PS-PEO diblock
copolymer used in the experiments
\cite{Bhargava-Zheng-Li-Quirk-Harris-Cheng-2006} is typically
$\simeq 1000$, thus we expect that $\chi_{\text{PS,PEO}} N$ is
sufficiently larger than the critical value $\chi_{c} N = 10.495$
\cite{Matsen-Bates-1996}.
Note that the polyelectrolytes such as poly(acrylic acid) (PAA)
is often used as the hydrophilic subchian
\cite{Shen-Eisenberg-1999,Shen-Eisenberg-2000}.
Such polyelectrolytes can be dissolved into water easily, and thus we
expect the $\chi$ parameter between the polyelectrolytes and the water is
negative.

Thus we can say the diblock copolymers are in strong segregation region, and
the effect of water addition is expected
to be large for hydrophobic subchains, but not so large for hydrophilic
subchains. Thus we consider that our simulation is qualitatively
consistent with these experiments.
Note that, however, we cannot compare
the $\chi$ parameter used in the simulations and measured by experiments
directly because the value of the $\chi$ parameter depends on the
definition of the segment, and our free energy functional
\eqref{freeenergy_diblocksolution} is not quantitatively accurate as the SCF
theory \cite{Uneyama-Doi-2005}.
We also note that
most experiments are carried out in the very strong segregation
region in which it is quite hard to perform continuum field simulations.
While it is difficult to compare our simulations with real experiments,
our simulations are expected to give qualitatively correct
physical process since our simulations take the crucial physical
properties correctly; the hydrophobic interaction $\chi_{BS}$ is
sufficiently large and the hydrophilic interaction $\chi_{AS}$ is
small.
These properties are not well
considered in most of previous dynamics simulations.

Other interesting properties are the size of the vesicles, the content
of solvents inside vesicles, or the higher order structures.
These properties cannot be discussed from
our current results, since the system size is not large. We have only
one vesicle in the simulation box (Figures
\ref{vesicle_formation_snapshots},\ref{vesicle_formation_snapshots2} and
\ref{vesicle_formation_snapshots3}) and
there may be the finite size effect (see also Appendix \ref{finite_size_effect}).
Nevertheless, we believe our model and simulation results are still
valuable. These properties of vesicles should be observed if we perform
the simulations for large systems while it is too difficult to perform large
scale simulations by the current models, algorithms and CPU power.

The characteristic time scales are also interesting and important property. As
mentioned above, the real time scale can be estimated from the diffusion
data. Here we estimate the time scale by
using the empirical equation for diffusion coefficient of short
polystyrene melts (at $T_{g} + 125 {}^{\circ}\text{C}$) by Watanabe and Kotaka \cite{Watanabe-Kotaka-1986}.
\begin{equation}
 D \simeq 6.3 \times 10^{-5} M^{-1} \mathrm{cm^{2} / s}
\end{equation}
where $M$ is the molecular weight.

We have to determine the number of monomers in one segment used in our
simulations. Since the polymerization index of typical amphiphilic block
copolymers used in experiments
\cite{Shen-Eisenberg-2000,Bhargava-Zheng-Li-Quirk-Harris-Cheng-2006} are
about $M \simeq 50 \sim 1000$, we consider that each segment used in
simulations contains about $10$ monomer units. In our simulations, we
used dimensionless segment size ($b = 1$). Thus we also need the size of the segment.
The segment size of the styrene monomer is about $7 \textrm{\AA}$
(it can be calculated from the experimental data of the radius of
gyration\cite{PoLyInfo,Wignall-Ballard-Schelten-1974}).
So if we assume the Gaussian
statistics for monomers in the segment, size of the segment can be
expressed as $b_{0} \simeq \sqrt{10} \times 7 \textrm{\AA}$.

Now we can estimate the characteristic time scale for the system.
If we assume that the polymer chains are not entangled and obey the Rouse dynamics,
the characteristic time scale $\tau$ can be calculated as follows.
\begin{equation}
 \tau = \frac{2 \zeta b_{0}^{2}}{k_{B} T} = \frac{2 b_{0}^{2}}{N D}
\end{equation}
$b_{0}$ is the size of the segment and $N$ is the number of segments in
one polymer chain (polymerization index), and $D$ is the diffusion coefficient.
$N$ and $M$ can be related as
$M = 104 \times 10 \times N$
and thus we have
\begin{equation}
 \tau \simeq \frac{2 \times 1040 \times \left(\sqrt{10} \times 7 \times 10^{-10} \mathrm{m} \right)^{2}}{6.5 \times 10^{-9} \mathrm{m}^{2} / \mathrm{s}}
\simeq 1.6 \times 10^{-6} \mathrm{s}
\end{equation}
Thus we know that the characteristic time scale in our simulations are
about $1$ microsecond, and the time scale of the vesicle formation is
about $10$ milliseconds.
Of course the time scale estimated here is quite rough and cannot be
compared with real experiments directly, but the
time scale of our simulation is considered to be much smaller than one
of experiments. The possible reasons are that the accuracy of the
dynamic equation used for the simulation is not so good, and that
the sizes of the formed vesicles are small since the system size is small.

\subsection{Comparison with the EPD simulations}
\label{comparison_with_epd_simulations}

We compare our simulation results with EPD simulation results and
discuss the reason why the EPD simulations
\cite{He-Schmid-2006} and our DF simulations give qualitatively
different results (the mechanism II and the mechanism I).

The main difference between the EPD simulations and our DF simulations
are the free energy functional model and the dynamic equation.
The EPD simulations use the free energy calculated from the SCF theory
while the DF simulations use the free energy functional
\eqref{freeenergy_diblocksolution}. It is known that the DF theory is
less accurate than the SCF theory, but gives qualitatively acceptable
results \cite{Uneyama-Doi-2005,Uneyama-Doi-2005a}. Thus it is difficult
to consider that the accuracy of the free energy model affect the
observed mechanisms. We expect that the mechanism I is reproduced if we
change the free energy functional model from eq
\eqref{freeenergy_diblocksolution} to more accurate SCF model.

Other difference is the dynamic equation. The EPD model employs nonlocal
mobility model, but in the derivation of the dynamic equation,
rather rough approximations are involved \cite{Muller-Schmid-2005}.
Because of these
approximations, the local mass conservation is not satisfied in the EPD
simulations  (unless the system is homogeneous). This may affect the dynamic behaviour seriously.
The mobility in our dynamic equation
\eqref{dynamic_density_functional_equation} is the local type, and thus
it is considered to be less accurate than the nonlocal model. But in our
dynamic equation, local mass conservation is satisfied exactly.
Thus we consider that the EPD simulations reproduce the mechanism II
because it does not satisfy the local mass conservation.
We expect that if the dynamic equation which satisfy the local mass conservation
is used with nonlocal mobility and the SCF free energy,
the mechanism I should be reproduced.

%------------------------------------------------------------------------------
\section{Conclusion}
\label{conclusion}
We have shown that we can reproduce the vesicle formation mechanism I
by the simulation using the dynamic density functional equation
\eqref{dynamic_density_functional_equation} and the free
energy functional \eqref{freeenergy_diblocksolution}.
The simulation results are qualitatively
in agreement with other simulations
\cite{Noguchi-Takasu-2001,Yamamoto-Maruyama-Hyodo-2002}.
This is the first realistic vesicle formation dynamics simulation based
on the field theoretical model. To reproduce the mechanism I, we need
the free energy functional which can be applied to strong segregation
regions, the large thermal noise, and the proper
interaction parameters for hydrophilic and hydrophobic interactions.
These conditions are not satisfied previous field theoretical dynamic
simulations.
We have also shown that the formation dynamics of various micellar
structures (spherical micelles,
cylindrical micelles, vesicles, bilayers) and morphological transitions
can be simulated by our model. The morphological transitions can be
reproduced only by changing the $\chi$ parameter which corresponds to
the change of the solvent condition.

The merit of using the field theoretical approach is that we can use the
Flory-Huggins interaction parameters, instead of potentials
between segments which is required for particle simulations.
We have shown that we can mimic the
solvent condition control by changing the $\chi$ parameter between
the hydrophobic segment and the solvent and have shown that the solvent
condition control actually causes morphological transitions.
For coarse grained particle simulations, we cannot use
the microscopic potential directly, and in most cases the
phenomenological potential models are used. Of course
for some particle simulations, such as the DPD, we can use the
$\chi$ parameters by mapping them onto the DPD potential parameters.
But such method is based on the parameter fitting \cite{Groot-Warren-1997}
and not always justified. Thus we consider that especially for
many component systems,
such as solutions of amphiphiles, the field theoretical approaches have
advantages.

Although the $\chi$ parameters
in the current model cannot be compared with the experiments directly,
our data may help understanding experimental data or the physical mechanism.
It will be possible to use the
experimentally determined $\chi$ parameters
by employing more
accurate free energy model such as the SCF model 
\cite{Fraaije-1993,Fredrickson-Ganesan-Drolet-2002,Kawakatsu-book}
and efficient numerical
algorithms. (We note that recently Honda and Kawakatsu
\cite{Honda-Kawakatsu-2006} proposed the
hybrid theory of the DF and the SCF and reported that 
accurate and numerically efficient simulations can be performed by the
hybrid method.
By employing their method, 
accurate and fast simulations for the dynamics of
micelles and vesicles may be possible.)
The quantitative simulations will be the future work.

While the efficiency of our simulation method based on the density functional theory
(eqs \eqref{freeenergy_diblocksolution} and
\eqref{dynamic_density_functional_equation_psi_rescaled}) is larger than
the SCF simulations, currently it is still comparable to particle simulations.
Besides, because the late stage of
the mechanism I is a very slow process, the current model
still need considerable computational costs.
To study the late stage dynamics or large scale systems (for example,
many vesicle systems), we will need more coarse
grained and numerically more efficient computational methods.
Our current work can be the basis of the further coarse grained field
theoretical model.

%------------------------------------------------------------------------------
\section{Acknowledgement}
\label{acknowledgment}
The author thanks T. Ohta 
for reading the manuscript and giving useful suggestions,
and for helpful comments and discussions.
Thanks are also due to H. Morita, T. Taniguchi, Y. Masubuchi, H. Frusawa, and
T. Kawakatsu for helpful comments and discussions.
The author also thanks anonymous referees who read the original manuscript and
gave him useful comments for the improvement of the manuscript.
This work is supported by the Research Fellowships of the Japan Society for
the Promotion of Science for Young Scientists.

%------------------------------------------------------------------------------
% appendix
\appendix

\section*{Appendix}

%------------------------------------------------------------------------------
\section{Noise Generation Scheme}
\label{noise_generation_scheme}
In this section we derive the numerical scheme for generating random noise
field which satisfies the fluctuation dissipation relation
(eqs \eqref{fluctuation_dissipation_relation_rescaled_first} and 
\eqref{fluctuation_dissipation_relation_rescaled_second}).
Such a noise field can be expressed as \cite{vanVlimmeren-Fraaije-1996}
\begin{equation}
 \label{tilde_xi_explicit}
 \begin{split}
  \tilde{\xi}_{i}(\bm{r},\tilde{t}) & = \sqrt{4 \tilde{\beta}^{-1}} \nabla \cdot \left[ \sqrt{\phi_{i}(\bm{r})} \bm{\omega}_{i}(\bm{r},\tilde{t}) \right] \\
  & = 2 \sqrt{\tilde{\beta}^{-1}} \nabla \cdot \left[ \psi_{i}(\bm{r}) \bm{\omega}_{i}(\bm{r},\tilde{t}) \right]
 \end{split}
\end{equation}
where $\bm{\omega}_{i}(\bm{r},\tilde{t})$ is the Gaussian white noise
(vector) field which satisfies
\begin{align}
 \label{fluctuation_dissipation_relation_omega_first}
 \left\langle \bm{\omega}_{i}(\bm{r},\tilde{t}) \right\rangle & = 0 \\
 \label{fluctuation_dissipation_relation_omega_second}
 \left\langle \bm{\omega}_{i}(\bm{r},\tilde{t}) \bm{\omega}_{j}(\bm{r}',\tilde{t}') \right\rangle & =
 \delta_{ij} \delta(\bm{r} - \bm{r}') \delta(\tilde{t} - \tilde{t}') \bm{1}
\end{align}
where $\bm{1}$ is the unit tensor.

To generate the noise numerically, we need the descretized version of eq
\eqref{tilde_xi_explicit}. The descretized noise field
$\tilde{\xi}_{i}(\bm{r},\tilde{t})$ is defined only on the lattice points.
Here we express the position of the lattice point as $\bm{r} = n_{x} \bm{h}_{x} + n_{y} \bm{h}_{y} + n_{z}
\bm{h}_{z}$ where $\bm{h}_{\alpha}$ ($\alpha = x,y,z$) is the lattice vector and
$n_{\alpha}$ is the integer.
The rescaled time $\tilde{t}$ is also
descretized as $\tilde{t} = n_{t} \Delta\tilde{t}$ where $n_{t}$ is the
integer.
The Laplacian operator is replaced by the standard centre finite difference
operator as follows
\begin{equation}
 \label{descretized_laplacian_definition}
 \nabla^{2} f(\bm{r}) \to \sum_{\alpha = x,y,z} \frac{1}{|\bm{h}_{\alpha}|^{2}} \left[ f(\bm{r} + \bm{h}_{\alpha}) - f(\bm{r}) + f(\bm{r} - \bm{h}_{\alpha}) \right]
\end{equation}
where $f(\bm{r})$ is a descretized field defined on the lattice points.
The delta functions are replaced as follows
\begin{align}
 \label{descretized_delta_r_definition}
 \delta(\bm{r} - \bm{r}') & \to
 \frac{\delta_{\bm{r} \bm{r}'}}{|\bm{h}_{x}| |\bm{h}_{y}| |\bm{h}_{z}|} \\
 \label{descretized_delta_t_definition}
 \delta(t - t') & \to \frac{\delta_{t t'}}{\Delta\tilde{t}}
\end{align}
where $\delta_{\bm{r} \bm{r}'} = \delta_{n_{x} n'_{x}} \delta_{n_{y}
n'_{y}} \delta_{n_{z} n'_{z}}$ and $\delta_{\tilde{t} \tilde{t}'} =
\delta_{n_{t} n'_{t}}$.

The fluctuation dissipation relation
for the second order moment (eq \eqref{fluctuation_dissipation_relation_rescaled_second}) 
can be rewritten as
\begin{equation}
 \label{fluctuation_dissipation_relation_rescaled_second_modified}
  \begin{split}
   \left\langle \tilde{\xi}_{i}(\bm{r},\tilde{t}) \tilde{\xi}_{j}(\bm{r}',\tilde{t}') \right\rangle 
   & = - 4 \tilde{\beta}^{-1} \delta_{ij} \nabla \cdot
   \left[ \psi^{2}_{i}(\bm{r}) \nabla \delta(\bm{r} - \bm{r}') \right]
   \delta(\tilde{t} - \tilde{t}') \\
   & = - 4 \tilde{\beta}^{-1} \delta_{ij} \nabla \cdot
   \left[ \psi^{2}_{i}(\bm{r}) \nabla \frac{\psi_{i}(\bm{r}) \delta(\bm{r} - \bm{r}')}{\psi_{i}(\bm{r})} \right]
   \delta(\tilde{t} - \tilde{t}') \\
   & = - 4 \tilde{\beta}^{-1} \delta_{ij} 
   \left[ \psi_{i}(\bm{r}) \nabla^{2} \left[ \psi_{i}(\bm{r}) \delta(\bm{r} - \bm{r}') \right] 
          - \left[ \psi_{i}(\bm{r}) \delta(\bm{r} - \bm{r}') \right] \nabla^{2} \psi_{i}(\bm{r}) \right]
   \delta(\tilde{t} - \tilde{t}')
  \end{split}
\end{equation}
Using eqs \eqref{descretized_laplacian_definition} -
\eqref{descretized_delta_t_definition}, we get the descretized version of the fluctuation dissipation relation
\eqref{fluctuation_dissipation_relation_rescaled_second_modified}.
\begin{equation}
 \label{fluctuation_dissipation_relation_rescaled_second_descretized}
  \begin{split}
   \left\langle \tilde{\xi}_{i}(\bm{r},\tilde{t}) \tilde{\xi}_{j}(\bm{r}',\tilde{t}') \right\rangle
   \approx - \frac{4 \tilde{\beta}^{-1} \delta_{ij}}{|\bm{h}_{x}| |\bm{h}_{y}| |\bm{h}_{z}| \Delta\tilde{t}}
   \sum_{\alpha = x,y,z} & \frac{1}{|\bm{h}_{\alpha}|^{2}}
   \bigg[ \psi_{i}(\bm{r} + \bm{h}_{\alpha}) \psi_{i}(\bm{r}) \left[ \delta_{(\bm{r} + \bm{h}_{\alpha}) \bm{r}'} - \delta_{\bm{r} \bm{r}'} \right] \\
   & - \psi_{i}(\bm{r}) \psi_{i}(\bm{r} - \bm{h}_{\alpha}) \left[ \delta_{\bm{r} \bm{r}'} - \delta_{(\bm{r} - \bm{h}_{\alpha}) \bm{r}'} \right] \bigg]
   \delta_{\tilde{t} \tilde{t}'}
  \end{split}
\end{equation}

In this work, we employ the following descretized equation to generate the noise field.
\begin{equation}
 \label{tilde_xi_descretized}
  \begin{split}
   \tilde{\xi}_{i}(\bm{r},\tilde{t}) \approx 2 \sqrt{\tilde{\beta}^{-1}} \sum_{\alpha = x,y,z} \frac{1}{|\bm{h}_{\alpha}|} \Big[ & \sqrt{\psi_{i}(\bm{r} + \bm{h}_{\alpha}) \psi_{i}(\bm{r})} \, \tilde{{\omega}}^{(\alpha)}_{i}(\bm{r} + \bm{h}_{\alpha} / 2,\tilde{t}) \\
   & - \sqrt{\psi_{i}(\bm{r}) \psi_{i}(\bm{r} - \bm{h}_{\alpha})} \, \tilde{{\omega}}^{(\alpha)}_{i}(\bm{r} - \bm{h}_{\alpha} / 2,\tilde{t}) \Big]
  \end{split}
\end{equation}
where $\tilde{\omega}^{(\alpha)}_{i}(\bm{r},\tilde{t})$ is the $\alpha$
element of $\tilde{\bm{\omega}}_{i}(\bm{r},\tilde{t})$.
$\tilde{\bm{\omega}}_{i}(\bm{r},\tilde{t})$ is the descretized
Gaussian white noise field which satisfies
\begin{align}
 \label{fluctuation_dissipation_relation_omega_descretized_first}
 \left\langle \tilde{\bm{\omega}}_{i}(\bm{r},\tilde{t}) \right\rangle & = 0 \\
 \label{fluctuation_dissipation_relation_omega_descretized_second}
 \left\langle \tilde{\bm{\omega}}_{i}(\bm{r},\tilde{t}) \tilde{\bm{\omega}}_{j}(\bm{r}',\tilde{t}') \right\rangle & =
 \frac{\delta_{ij} \delta_{\bm{r} \bm{r}'} \delta_{\tilde{t} \tilde{t}'}}{|\bm{h}_{x}| |\bm{h}_{y}| |\bm{h}_{z}| \Delta\tilde{t}} \bm{1}
\end{align}
Notice that $\tilde{\bm{\omega}}_{i}(\bm{r},\tilde{t})$ is defined on the
staggered lattice. The position of the staggered lattice point is
$\bm{r} = (n_{x} + 1/2) \bm{h}_{x} + n_{y} \bm{h}_{y} + n_{z} \bm{h}_{z},
n_{x} \bm{h}_{x} + (n_{y} + 1/2) \bm{h}_{y} + n_{z} \bm{h}_{z}, 
n_{x} \bm{h}_{x} + n_{y} \bm{h}_{y} + (n_{z} + 1/2) \bm{h}_{z}$
where $n_{\alpha}$ is the integer.
It is easy to show that the noise field generated by eq
\eqref{tilde_xi_descretized} satisfies eq
\eqref{fluctuation_dissipation_relation_rescaled_second_descretized}.

The noise generation scheme is finally written as follows.
\begin{enumerate}
 \item Generate the Gaussian white normal distribution vector noise
       field $\tilde{\bm{\omega}}_{i}(\bm{r},\tilde{t})$ on the staggered lattice
       points.
       The noise is generated by using the Mersenne twister
       pseudo-random number generator \cite{Matsumoto-Nishimura-1998}
       and the standard Box-Muller method \cite{NumericalRecipes}.
 \item Calculate the noise field $\tilde{\xi}_{i}(\bm{r},\tilde{t})$
       from $\psi_{i}(\bm{r})$ and $\tilde{\bm{\omega}}_{i}(\bm{r},\tilde{t})$ by
       using eq \eqref{tilde_xi_descretized}.
\end{enumerate}

%------------------------------------------------------------------------------
\section{Finite Size Effect}
\label{finite_size_effect}
It is well known that the size of the simulation box often affects the
simulation results (the finite size effect). In this appendix, we study
the finite size effect
for our simulations by performing simulations with different box size.

Figure \ref{small_boxes} shows the simulation results for different
(small) box sizes. All the parameters except for the box size are the
same for the previous vesicle formation simulation (Figure
\ref{vesicle_formation_snapshots}). The box sizes are set to
$6b \times 6b \times 6b$ ($12 \times 12 \times 12$ lattice points),
$8b \times 8b \times 8b$ ($16 \times 16 \times 16$ lattice points),
$12b \times 12b \times 12b$ ($24 \times 24 \times 24$ lattice points),
and $16b \times 16b \times 16b$ ($32 \times 32 \times 32$ lattice points).
We can observe the spherical micelles (Figure
\ref{small_boxes}(a)) and the cylindrical micelles (Figure
\ref{small_boxes}(b),(c)) for small boxes. These structures are much
different from the
vesicle structures in Figure \ref{small_boxes}(d) or Figure \ref{vesicle_formation_snapshots}.
This means that if the simulation box is too small, we cannot obtain
vesicles.

It also implies that there may be the finite size effect
in our vesicle formation simulations (box size $48 \times 48 \times
48$), since the size of obtained vesicles are in most cases comparable
to the box sizes.
But at least we can claim that our vesicle formation process itself is
not affected qualitatively since the system size is not so small for the
formation of disk like structures (unlike the case of Figure
\ref{small_boxes}(a),(b),(c)) and we can observe that the small vesicle formation
process (Figure \ref{vesicle_formation_snapshots2}(d)) is qualitatively
same as one for larger vesicles.

%------------------------------------------------------------------------------
\section{Cutoff Length Dependency}
\label{cutoff_length_dependency}
The cutoff length $\lambda$ for the long range interaction term in eq
\eqref{conformational_entropy} is introduced rather intuitively, and
therefore the value of $\lambda$ for simulations should be considered
carefully. In this section, we perform the simulations with different
values of $\lambda$. Simulations are performed in two dimensions.
Parameters are as follows:
$N = 10$, $f_{A} = 1/3, f_{B} = 2/3$ $\lambda =
2,4,5,6,8$. $\bar{\phi}_{s} = 1 - \bar{\phi}_{p} = 0.8$.
$\chi_{AB} = 2.5, \chi_{AS} = -0.5, \chi_{BS} = 5$, $b = 1$.
The size of the
2 dimensional simulation box is $64b \times 64b$ 
($128 \times 128$ lattice points, periodic boundary condition).
The time step is ${\Delta}{\tilde t} = 0.0025$ (for $\lambda =
2,4,5$), $0.000625$ (for $\lambda = 6$), ${\Delta}{\tilde t} = 0.0003125$ (for $\lambda = 8$).
The magnitude of the noise is $\tilde{\beta}^{-1} =  0.078125$.
All the simulations are started from the homogeneous state.

Figure \ref{lambda_snapshots} shows the results for various value of the
cutoff length ((a),(b) $\lambda = 1$, (c),(d) $\lambda = 2$,
(e),(f) $\lambda = 4$, (g),(h) $\lambda = 5$, (i),(j) $\lambda = 6$,
and (k),(l) $\lambda = 8$)
at $\tilde{t} = 312.5, 1250$.
We can observe the droplet like structures are observed for $\lambda = 1$.
and micellar structures for $2 \le \lambda \le 8$.
The resulting morphologies for $2 \le \lambda \le 8$
are qualitatively same.
(Strictly speaking, $\lambda = 2$ looks slightly different from
others and somehow resembles to $\lambda = 1$.)
As expected, the cutoff length dependence of the morphology is
small. Thus even if the value of $\lambda$ is not accurate, the results
are expected not to be affected qualitatively.
From this result, we can justify to use the value $\lambda = 5$, which
is estimated by the rough argument.

%------------------------------------------------------------------------------
% references

%------------------------------------------------------------------------------
% figures

\clearpage

\section*{Figures}
\subsection*{Figure Captions}
\begin{figure}[p!]
 \centering
 {\includegraphics[width=0.95\linewidth,clip]{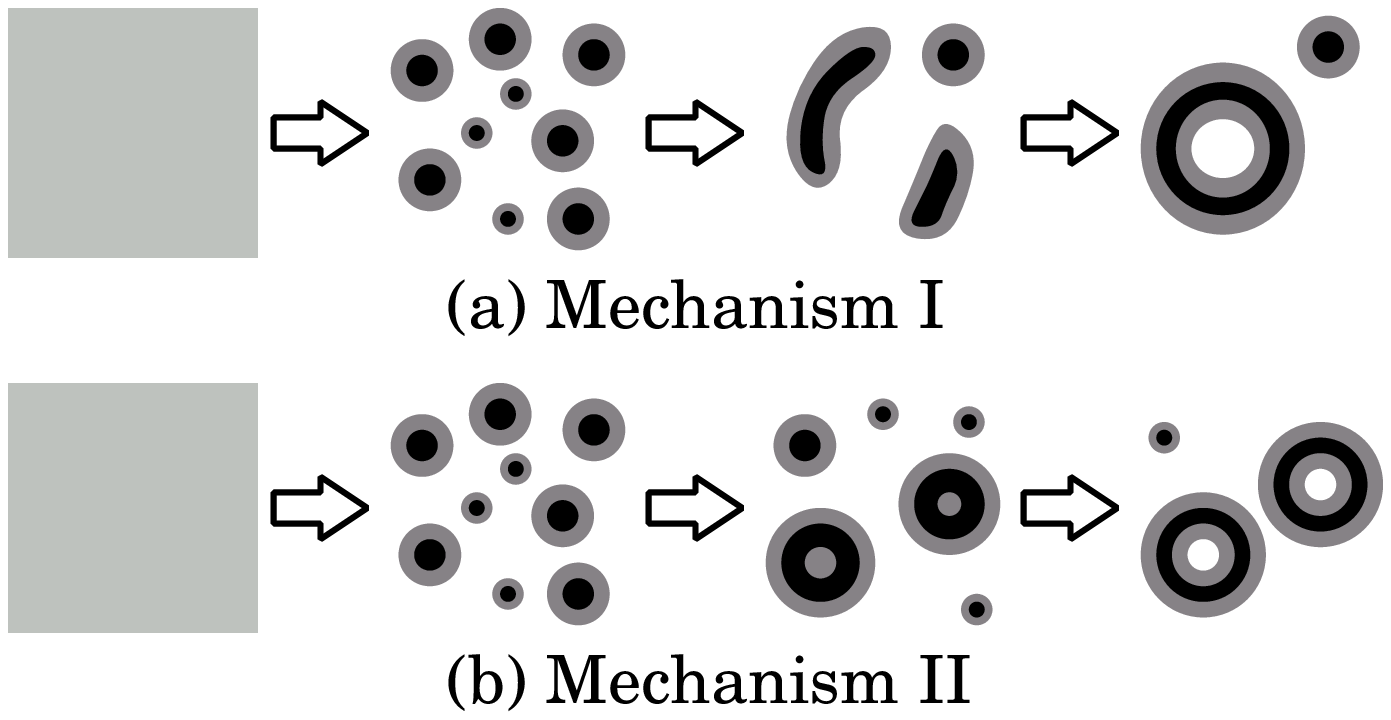}}
 \caption{}
 \label{vesicle_formation_mechanisms}
\end{figure}

\hspace{\parindent}%
Figure \ref{vesicle_formation_mechanisms}:
{Schematic representation of two vesicle formation mechanisms from the initial homogeneous
state. Black and grey color corresponds to hydrophobic and hydrophilic
subchains, respectively.
(a) mechanism I: First, small micellar structures are formed. The
micellar structures grow by collision and become large cylindrical or open disk-like
micelles. Finally the open disk-like micelle close up to form closed vesicles.
(b) mechanism II: The initial stage is similar to the mechanism I.
However, the small micellar structures formed in the initial stage
grows up to be large spherical micelles. The large spherical micelles
then take the solvent inside and thus vesicles are formed.}

\begin{figure}[p!]
 \centering
 {}
 {\includegraphics[width=0.45\linewidth,clip]{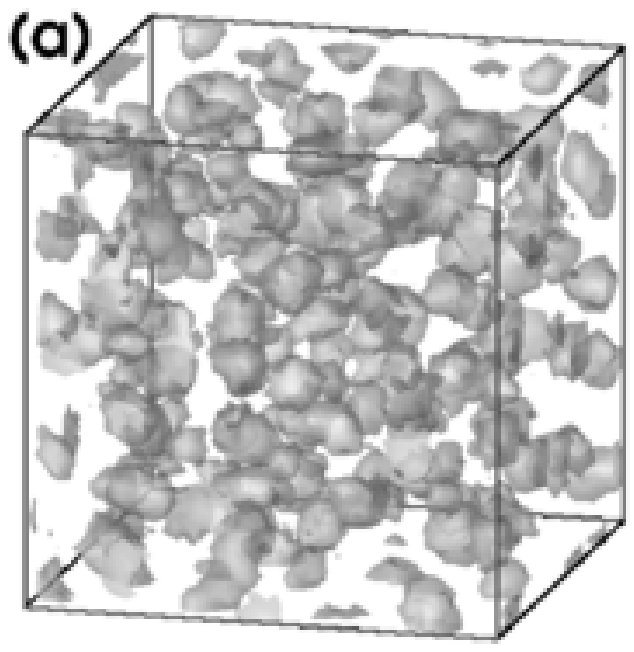}}
 \hspace{0.025\linewidth}
 {\includegraphics[width=0.45\linewidth,clip]{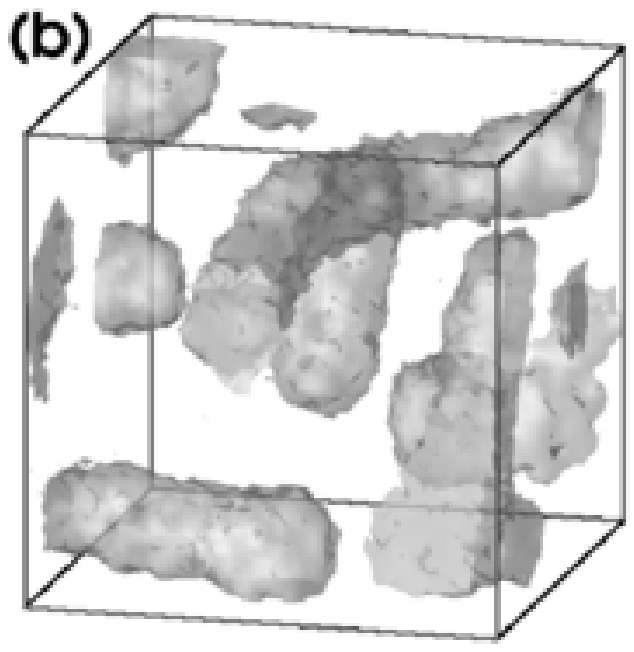}}
 \vspace{0.035\linewidth} {}
 {\includegraphics[width=0.45\linewidth,clip]{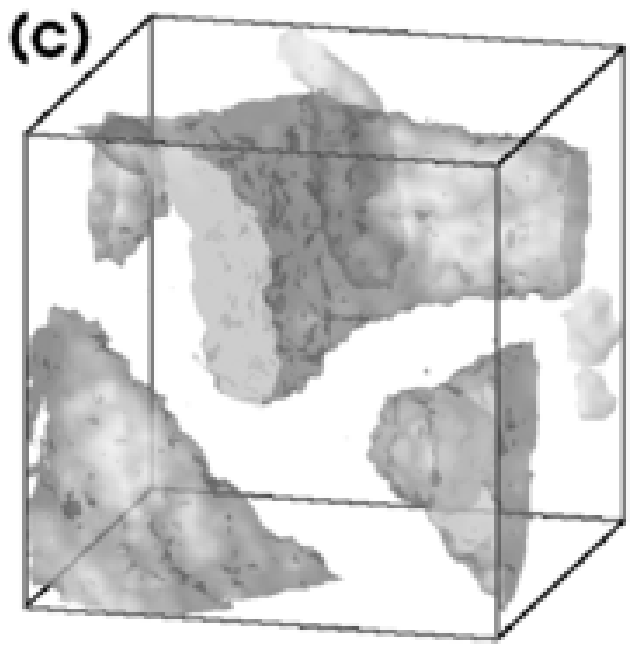}}
 \hspace{0.025\linewidth}
 {\includegraphics[width=0.45\linewidth,clip]{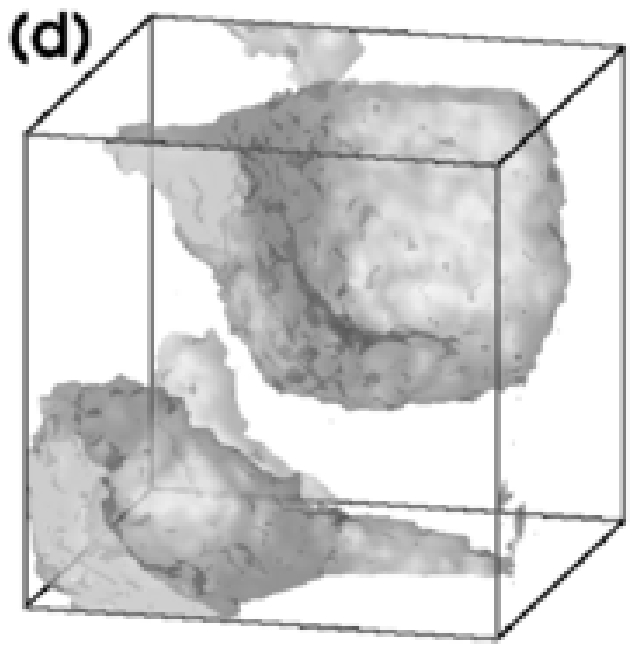}}
 \vspace{0.035\linewidth} {}
 {\includegraphics[width=0.45\linewidth,clip]{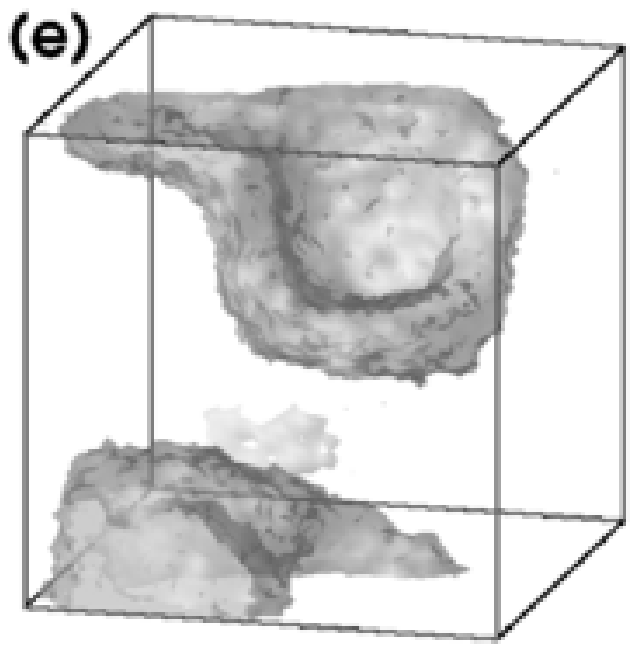}}
 \hspace{0.025\linewidth}
 {\includegraphics[width=0.45\linewidth,clip]{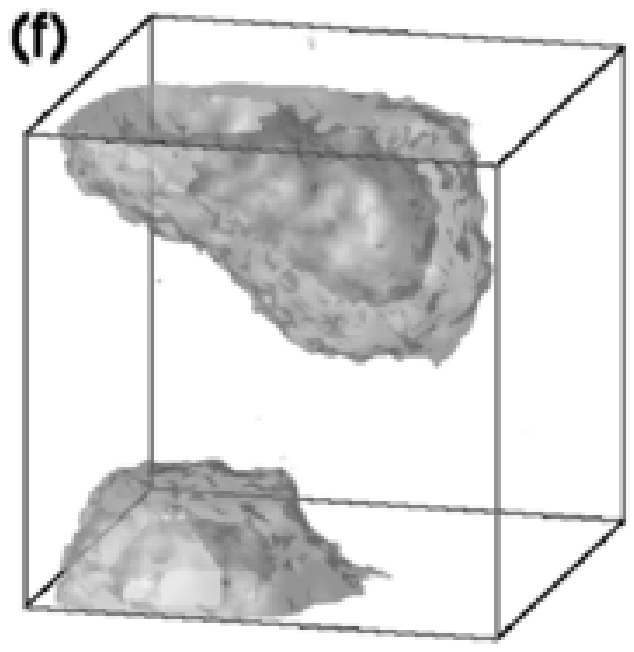}}
 {}
 \caption{}
 \label{vesicle_formation_snapshots}
\end{figure}

Figure \ref{vesicle_formation_snapshots}:
{Snapshots of the dynamics simulation for the amphiphilic diblock copolymer
solution ($\bar{\phi}_{p} = 0.2$). Grey surface is the isodensity
surface for the hydrophobic subchain, $\phi_{B}(\bm{r}) = 0.5$.
(a) $\tilde{t} = 6.25$, (b) $\tilde{t} = 312.5$, (c) $\tilde{t} = 1562.5$,
(d) $\tilde{t} = 3125$, (e) $\tilde{t} = 4687.5$, (f) $\tilde{t} = 6250$.}

\begin{figure}[p!]
 \centering
 {}
 {\includegraphics[width=0.45\linewidth,clip]{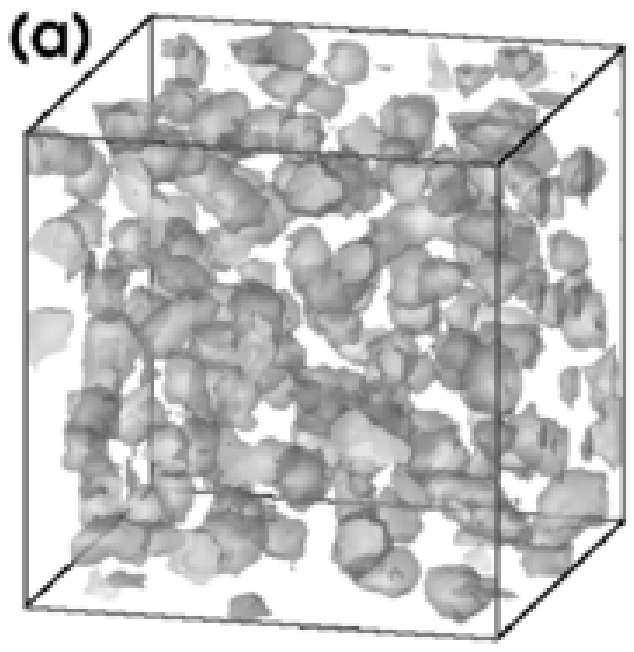}}
 \hspace{0.025\linewidth}
 {\includegraphics[width=0.45\linewidth,clip]{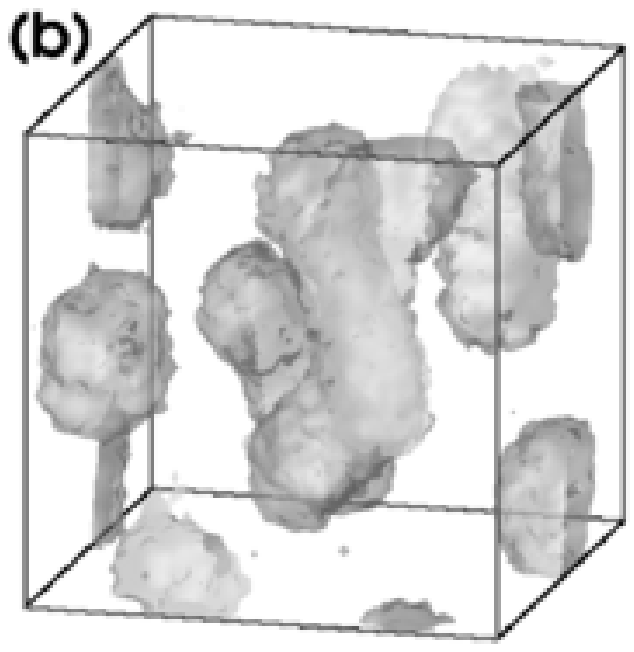}}
 \vspace{0.035\linewidth} {}
 {\includegraphics[width=0.45\linewidth,clip]{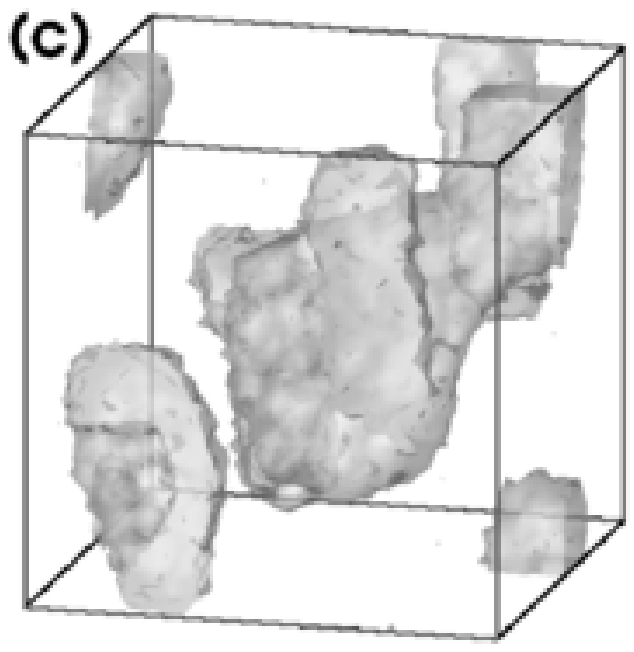}}
 \hspace{0.025\linewidth}
 {\includegraphics[width=0.45\linewidth,clip]{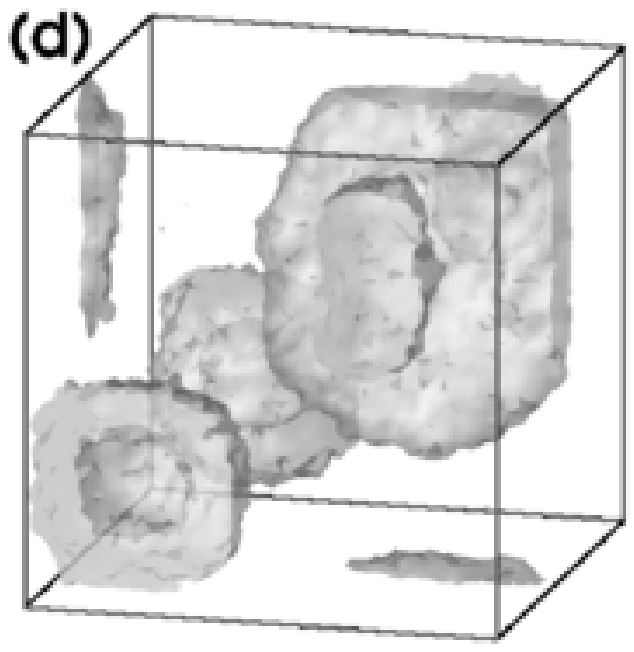}}
 {}
 \caption{}
 \label{vesicle_formation_snapshots2}
\end{figure}

Figure \ref{vesicle_formation_snapshots2}:
{Snapshots of the dynamics simulation for the amphiphilic diblock copolymer
solution ($\bar{\phi}_{p} = 0.2$). Grey surface is the isodensity
surface for the hydrophobic subchain, $\phi_{B}(\bm{r}) = 0.5$.
(a) $\tilde{t} = 6.25$, (b) $\tilde{t} = 312.5$, (c) $\tilde{t} = 1562.5$,
(d) $\tilde{t} = 3125$.}

\begin{figure}[p!]
 \centering
 {}
 {\includegraphics[width=0.45\linewidth,clip]{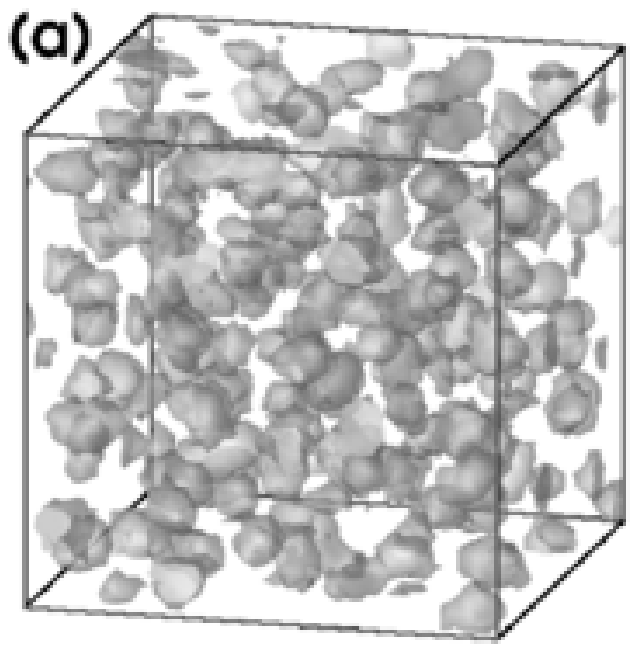}}
 \hspace{0.025\linewidth}
 {\includegraphics[width=0.45\linewidth,clip]{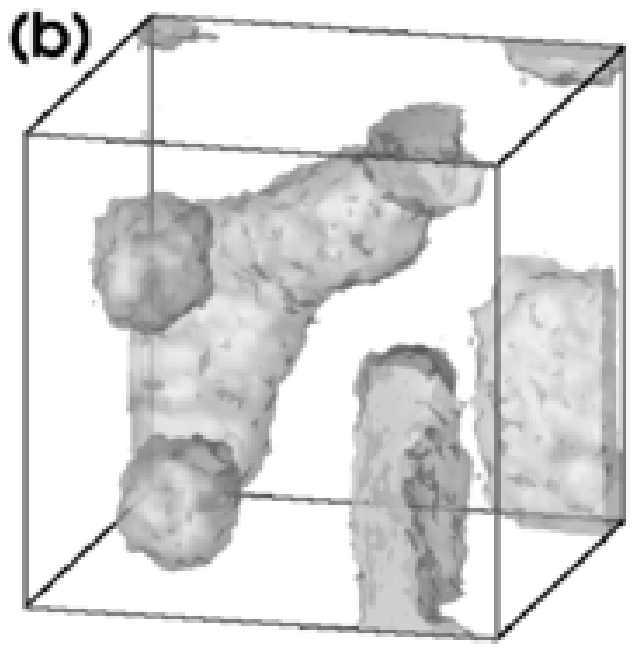}}
 \vspace{0.035\linewidth} {}
 {\includegraphics[width=0.45\linewidth,clip]{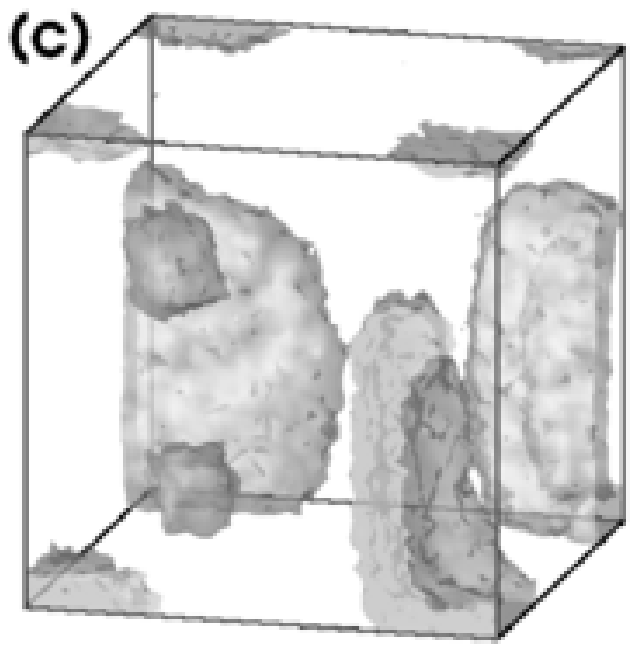}}
 \hspace{0.025\linewidth}
 {\includegraphics[width=0.45\linewidth,clip]{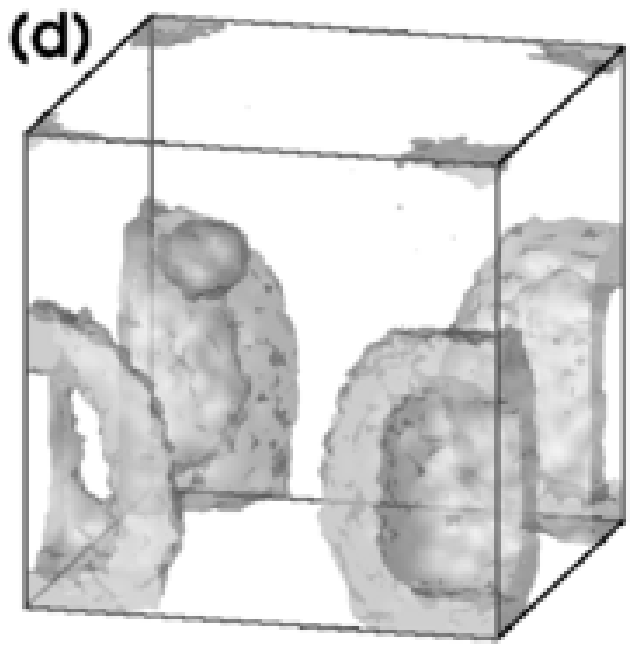}}
 {}
 \caption{}
 \label{vesicle_formation_snapshots3}
\end{figure}

Figure \ref{vesicle_formation_snapshots3}:
{Snapshots of the dynamics simulation for the amphiphilic diblock copolymer
solution ($\bar{\phi}_{p} = 0.2$). Grey surface is the isodensity
surface for the hydrophobic subchain, $\phi_{B}(\bm{r}) = 0.5$.
(a) $\tilde{t} = 6.25$, (b) $\tilde{t} = 3125$, (c) $\tilde{t} = 4687.5$,
(d) $\tilde{t} = 6250$.}

\begin{figure}[p!]
 \centering
 {\includegraphics[height=0.95\linewidth,clip,angle=-90]{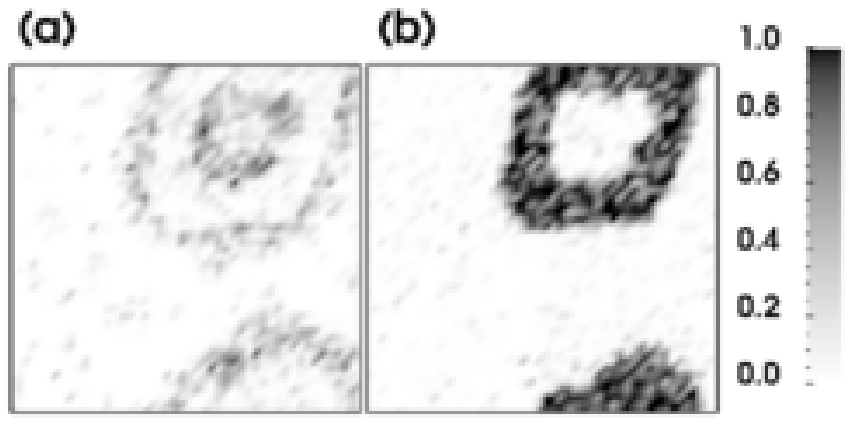}}
 \vspace{0.035\linewidth} {}
 {\includegraphics[width=0.8\linewidth,clip]{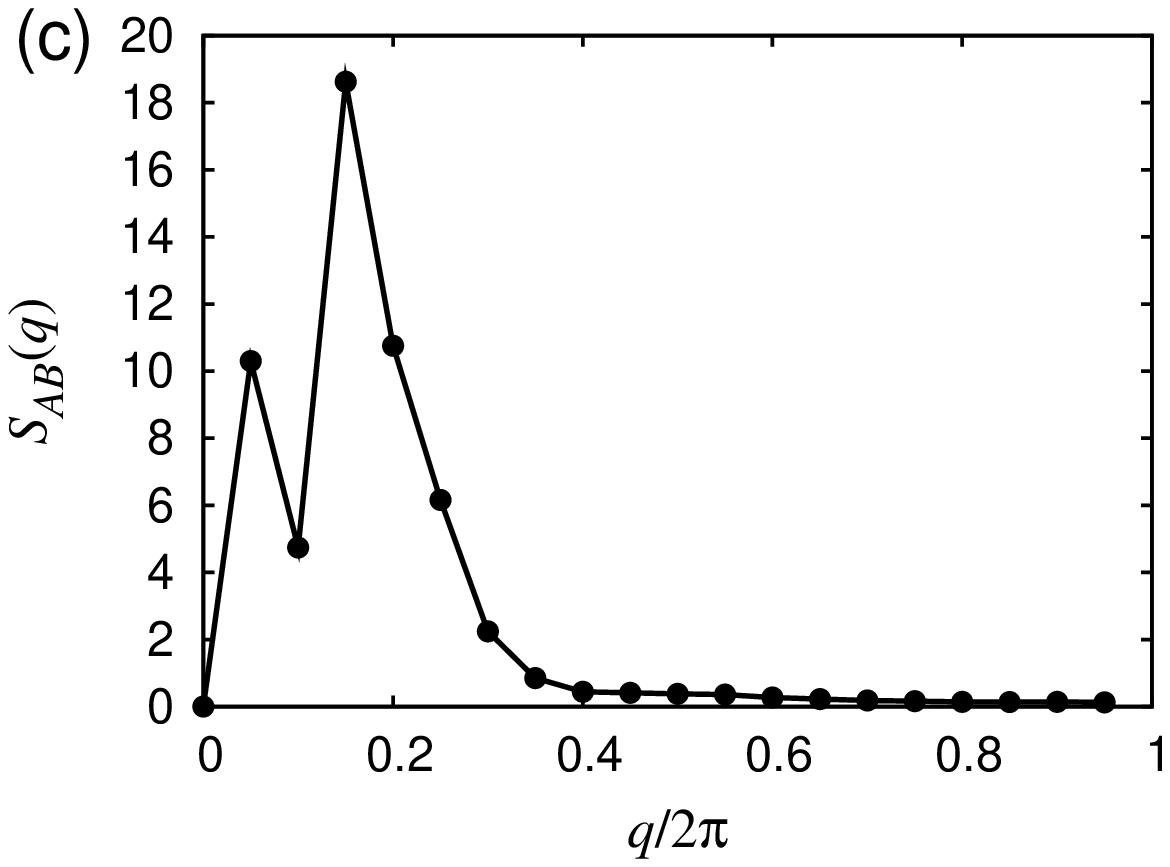}}
 \caption{}
 \label{cross_section_density_correlation}
\end{figure}

Figure \ref{cross_section_density_correlation}:
{Two dimensional cross sections of density field for (a) $\phi_{A}(\bm{r})$ and
(b) $\phi_{B}(\bm{r})$. (c) The density correlation function
$S_{AB}(\bm{q})$. $\tilde{t} = 6250$.}

\begin{figure}[p!]
 \centering
 {}
 {\includegraphics[width=0.45\linewidth,clip]{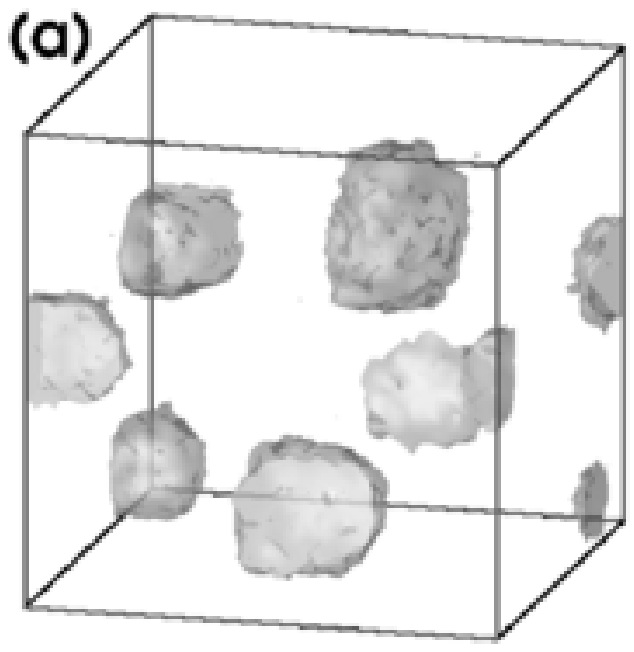}}
 \hspace{0.025\linewidth}
 {\includegraphics[width=0.45\linewidth,clip]{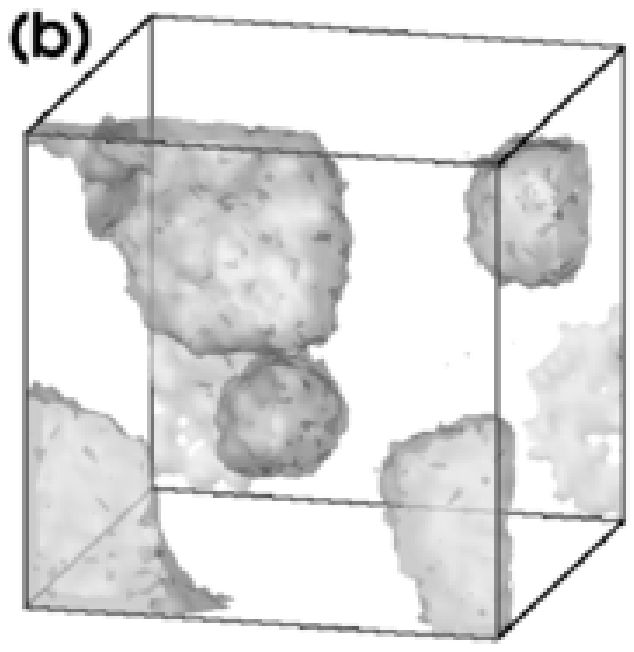}}
 \vspace{0.035\linewidth} {}
 {\includegraphics[width=0.45\linewidth,clip]{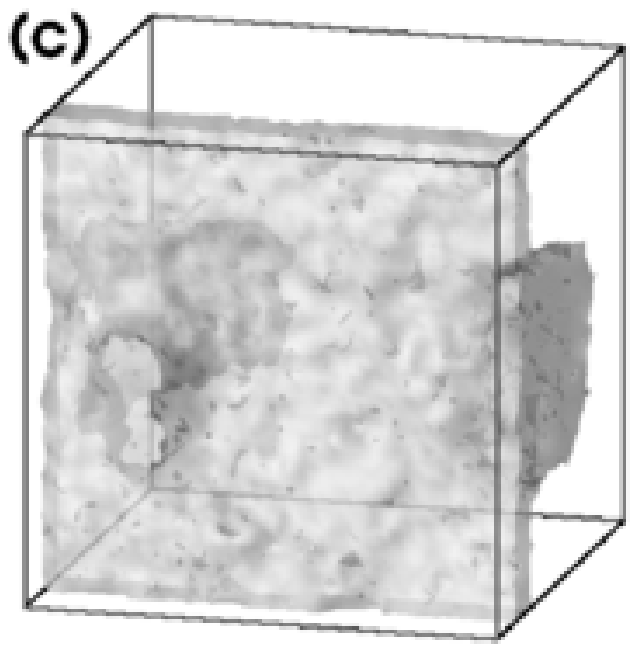}}
 \hspace{0.025\linewidth}
 {\includegraphics[width=0.45\linewidth,clip]{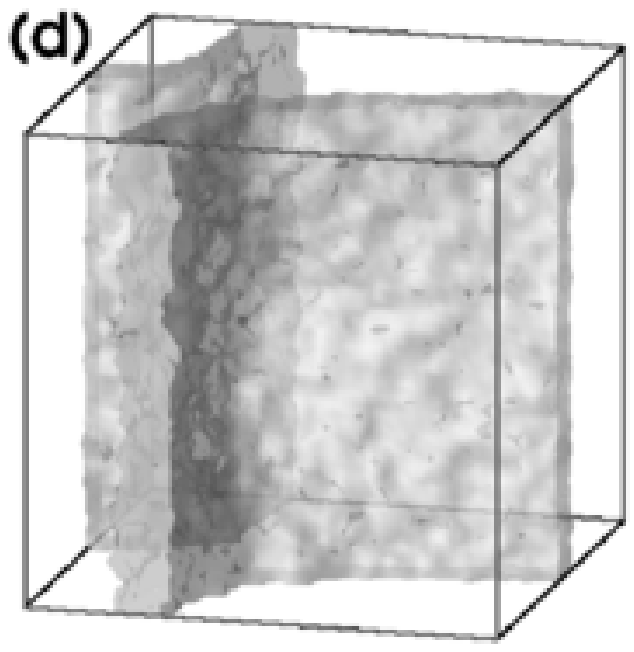}}
 {}
 \caption{}
 \label{structure_volume_fraction_dependence}
\end{figure}

Figure \ref{structure_volume_fraction_dependence}:
{Snapshots of the dynamics simulations for amphiphilic diblock copolymer
solutions with various volume fractions.
Grey surface is the isodensity surface for the hydrophobic subchain,
$\phi_{B}(\bm{r}) = 0.5$ at $\tilde{t} = 3125$.
(a) $\bar{\phi}_{p} = 0.1$, (b) $\bar{\phi}_{p} = 0.15$,
(c) $\bar{\phi}_{p} = 0.25$, (d) $\bar{\phi}_{p} = 0.3$.
($\bar{\phi}_{p} = 0.2$ corresponds to Figure \ref{vesicle_formation_snapshots}(d).)}

\begin{figure}[p!]
 \centering
 {}
 {\includegraphics[width=0.45\linewidth,clip]{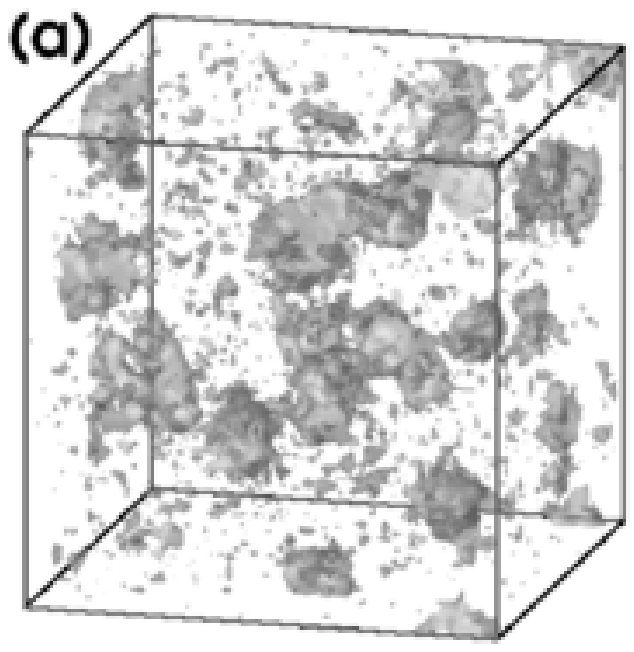}}
 \hspace{0.025\linewidth}
 {\includegraphics[width=0.45\linewidth,clip]{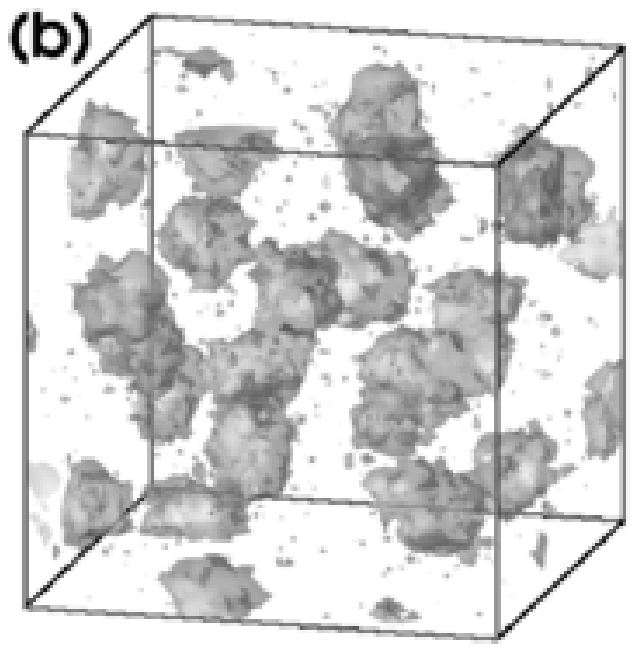}}
 \vspace{0.035\linewidth} {}
 {\includegraphics[width=0.45\linewidth,clip]{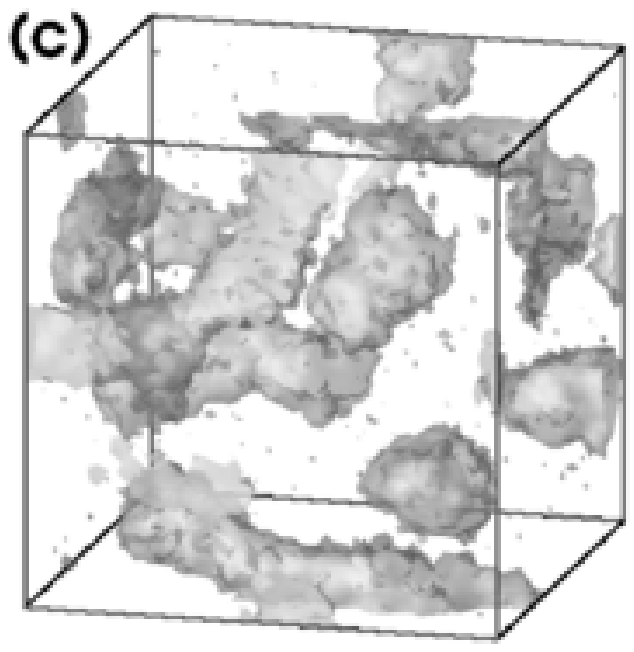}}
 \hspace{0.025\linewidth}
 {\includegraphics[width=0.45\linewidth,clip]{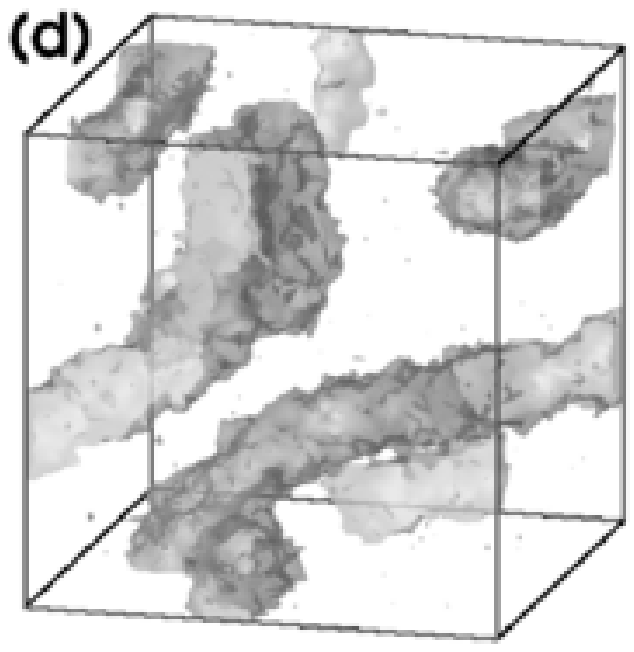}}
 {}
 \caption{}
 \label{structure_chi_parameter_dependence}
\end{figure}

Figure \ref{structure_chi_parameter_dependence}:
{Snapshots of the dynamics simulations for amphiphilic diblock copolymer
solutions with various $\chi_{BS}$ (the $\chi$ parameter between the
hydrophobic chain and the solvent).
Grey surface is the isodensity surface for the hydrophobic subchain,
$\phi_{B}(\bm{r}) = 0.5$ at $\tilde{t} = 3125$.
(a) $\chi_{BS} = 2$, (b) $\chi_{BS} = 2.5$,
(c) $\chi_{BS} = 3$, (d) $\chi_{BS} = 3.5$.
($\chi_{BS} = 5$ corresponds to Figure \ref{vesicle_formation_snapshots}(d).)}

\begin{figure}[p!]
 \centering
 {}
 {\includegraphics[width=0.45\linewidth,clip]{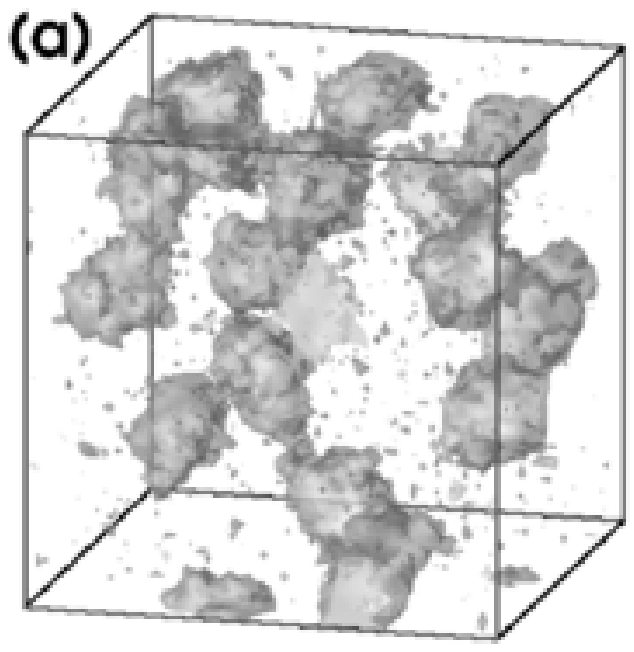}}
 \hspace{0.025\linewidth}
 {\includegraphics[width=0.45\linewidth,clip]{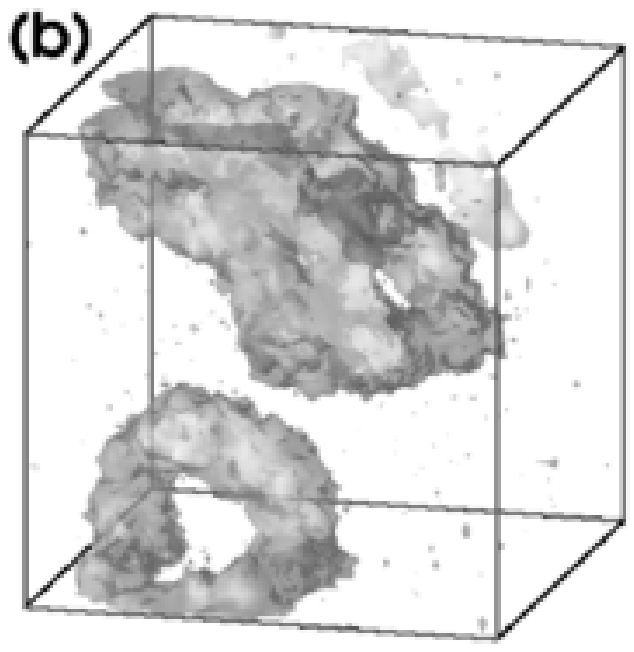}}
 {}
 \caption{}
 \label{structural_transition}
\end{figure}

Figure \ref{structural_transition}:
{Snapshots of the structural transition dynamics simulations. The $\chi$
parameter $\chi_{BS}$ is initially set to $\chi_{BS} = 5$ and at $\tilde{t} =
6250$, $\chi_{BS}$ is set to lower value.
Grey surface is the isodensity surface for the hydrophobic subchain,
$\phi_{B}(\bm{r}) = 0.5$ at $\tilde{t} = 9375$.
(a) $\chi_{BS} = 5 \to 2.5$, (b) $\chi_{BS} = 5 \to 3$.
}

\begin{figure}[p!]
 \centering
 {}
 {\includegraphics[width=0.45\linewidth,clip]{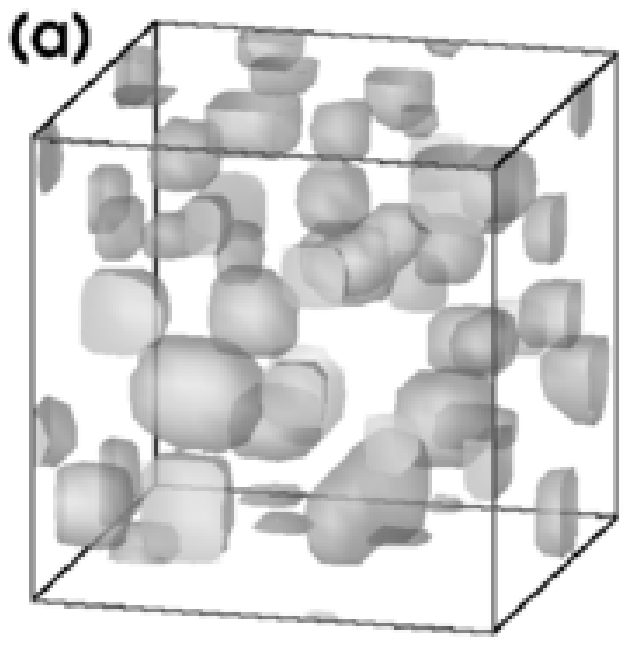}}
 \hspace{0.025\linewidth}
 {\includegraphics[width=0.45\linewidth,clip]{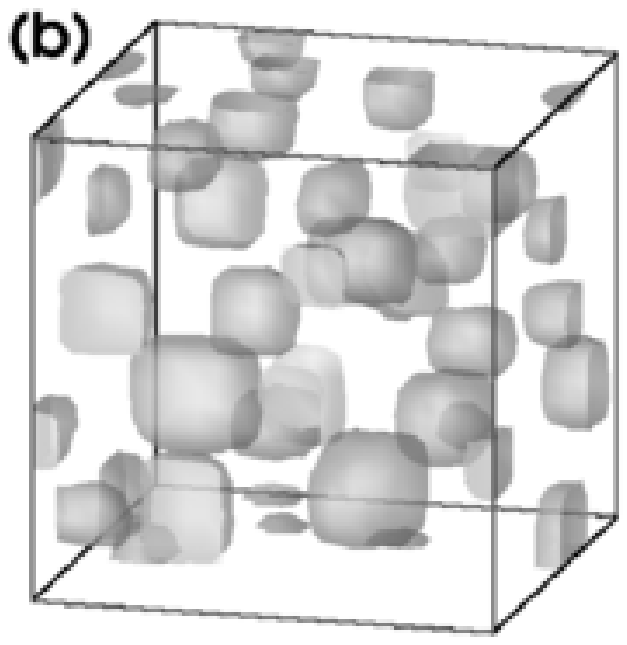}}
 \vspace{0.035\linewidth} {}
 {\includegraphics[width=0.45\linewidth,clip]{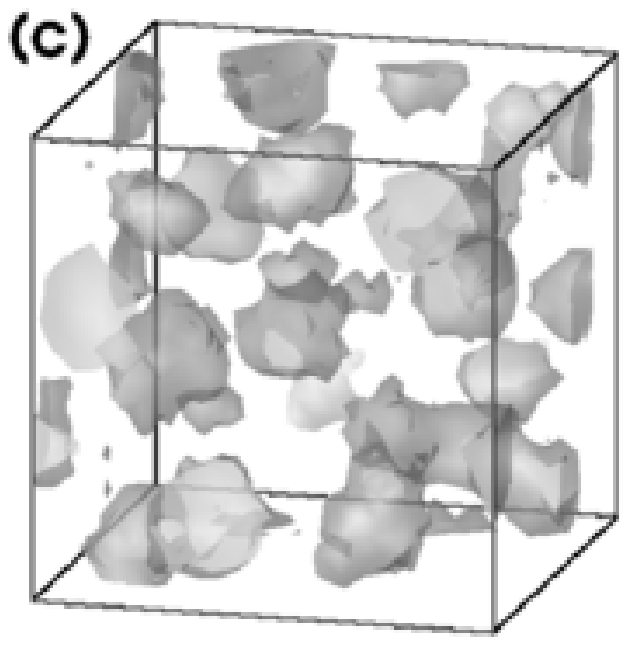}}
 \hspace{0.025\linewidth}
 {\includegraphics[width=0.45\linewidth,clip]{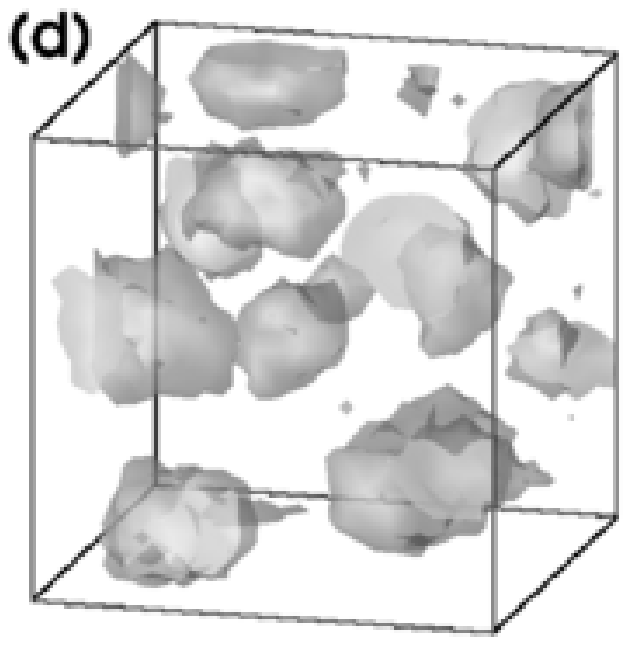}}
 \vspace{0.035\linewidth} {}
 {\includegraphics[width=0.45\linewidth,clip]{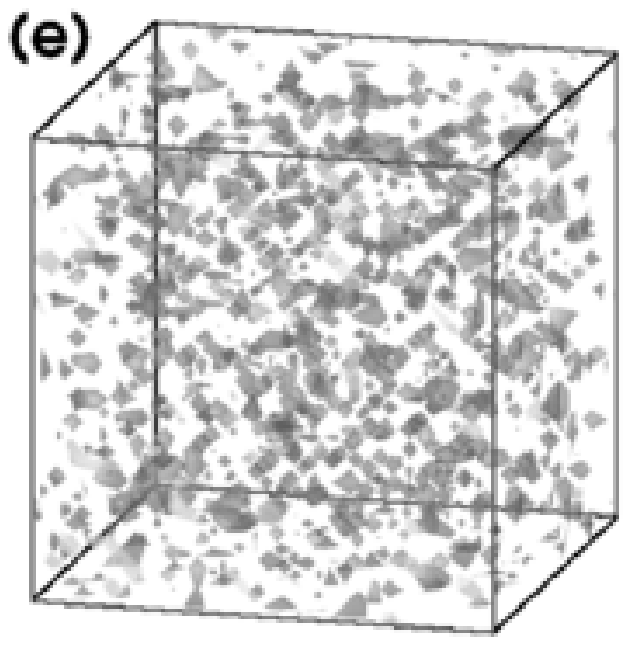}}
 \hspace{0.025\linewidth}
 {\includegraphics[width=0.45\linewidth,clip]{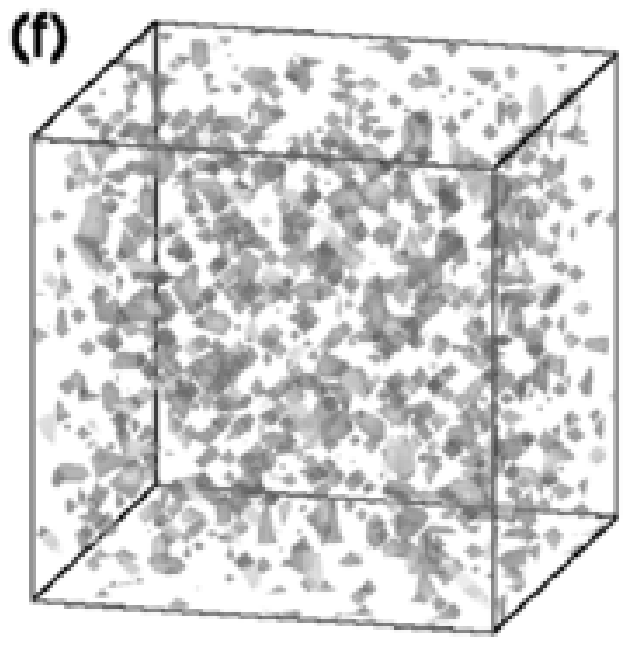}}
 {}
 \caption{}
 \label{thermal_noise_snapshots}
\end{figure}

Figure \ref{thermal_noise_snapshots}:
{Snapshots of the dynamics simulations for values of
$\tilde{\beta}^{-1}$ at $\tilde{t} = 6.25, 12.5$.
(a),(b) $\tilde{\beta}^{-1} = 0.00390625$, $\tilde{t} = 6.25, 12.5$.
(c),(d) $\tilde{\beta}^{-1} = 0.0390625$, $\tilde{t} = 6.25, 12.5$.
(e),(f) $\tilde{\beta}^{-1} = 0.390625$, $\tilde{t} = 6.25, 12.5$.
}

\begin{figure}[p!]
 \centering
 {}
 {\includegraphics[width=0.45\linewidth,clip]{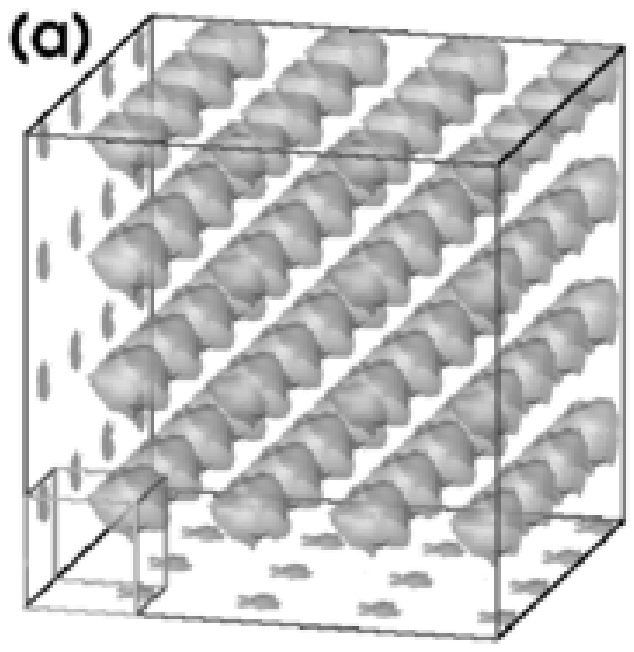}}
 \hspace{0.025\linewidth}
 {\includegraphics[width=0.45\linewidth,clip]{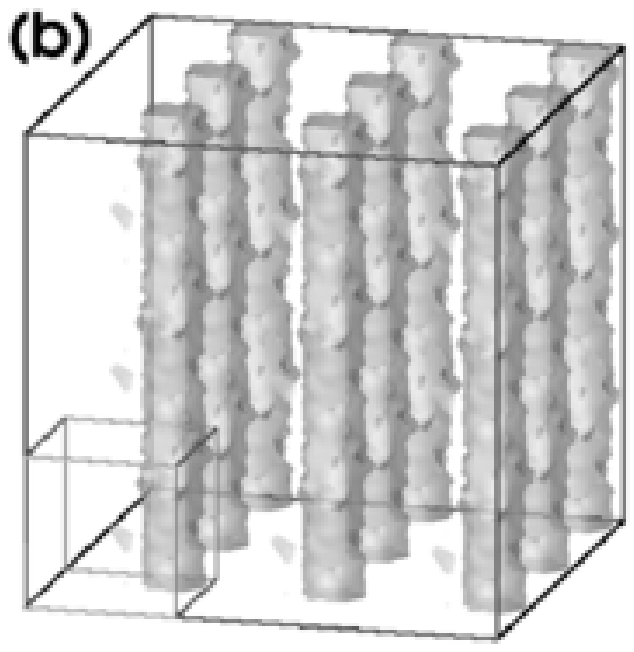}}
 \vspace{0.035\linewidth} {}
 {\includegraphics[width=0.45\linewidth,clip]{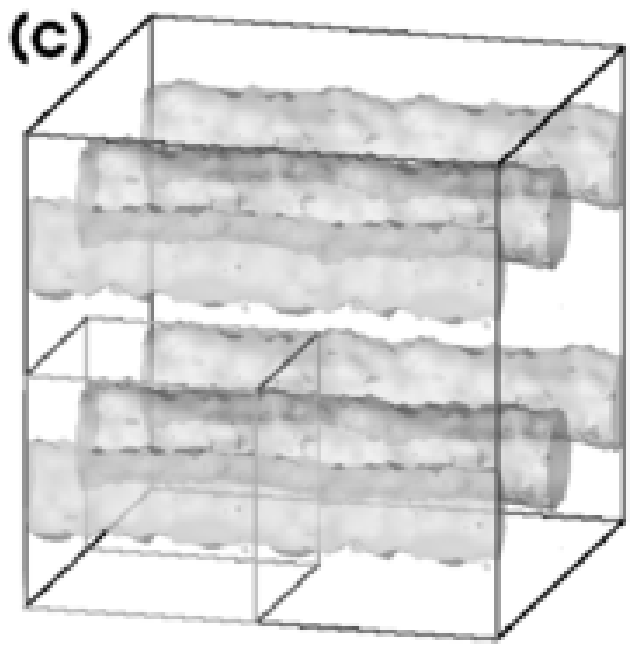}}
 \hspace{0.025\linewidth}
 {\includegraphics[width=0.45\linewidth,clip]{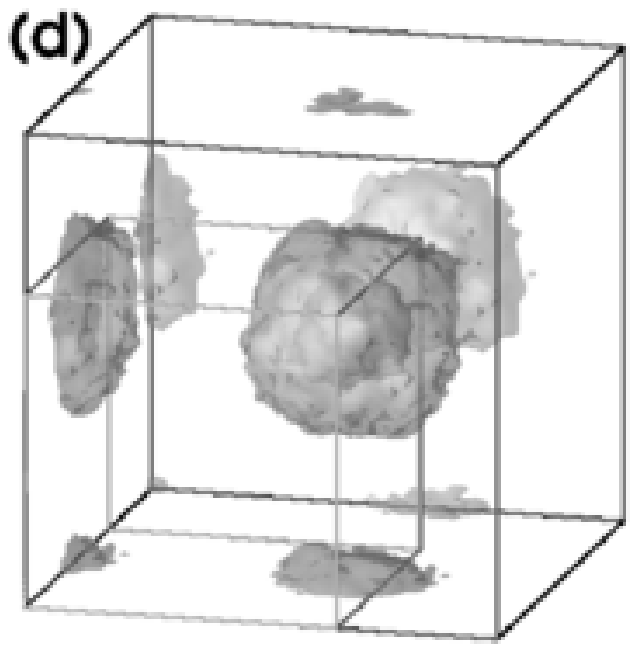}}
 {}
 \caption{}
 \label{small_boxes}
\end{figure}

Figure \ref{small_boxes}:
{Snapshots of the dynamics simulations for amphiphilic diblock copolymer
solutions with various small box sizes at $\tilde{t} = 3125$. (a) box
size $6b \times 6b \times 6b$ ($12 \times 12 \times 12$ lattice points),
(b) box size $8b \times 8b \times 8b$ ($16 \times 16 \times 16$ lattice points),
(c) box size $12b \times 12b \times 12b$ ($24 \times 24 \times 24$ lattice points),
and (d) box size $16b \times 16b \times 16b$ ($32 \times 32 \times 32$ lattice points).
Super cells (size $24b \times 24b \times 24b$) is shown. Small gray
boxes show the real simulation box.
}

\begin{figure}[p!]
 \centering
 {}
 {\includegraphics[width=0.24\linewidth,clip]{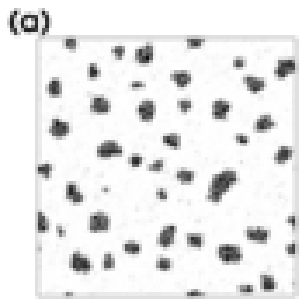}}
 {\includegraphics[width=0.24\linewidth,clip]{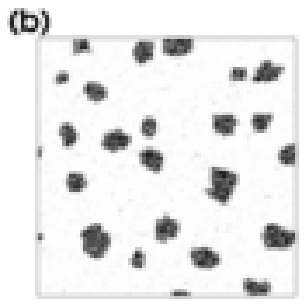}}
 {\includegraphics[width=0.24\linewidth,clip]{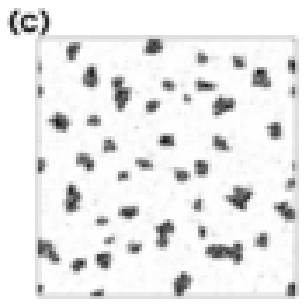}}
 {\includegraphics[width=0.24\linewidth,clip]{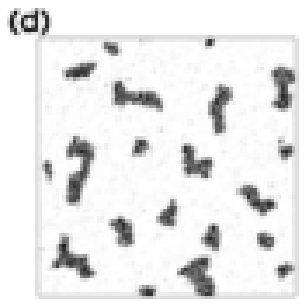}}
 \vspace{0.015\linewidth} {}
 {\includegraphics[width=0.24\linewidth,clip]{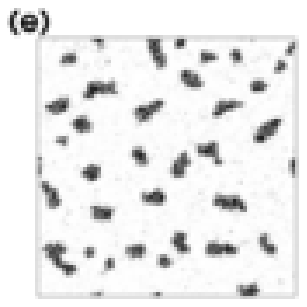}}
 {\includegraphics[width=0.24\linewidth,clip]{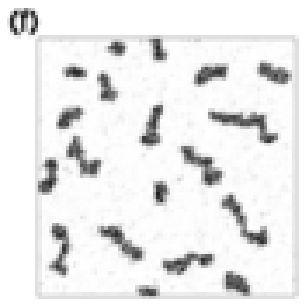}}
 {\includegraphics[width=0.24\linewidth,clip]{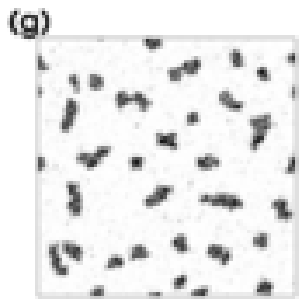}}
 {\includegraphics[width=0.24\linewidth,clip]{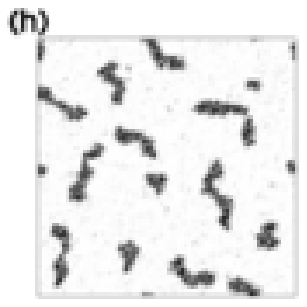}}
 \vspace{0.015\linewidth} {}
 {\includegraphics[width=0.24\linewidth,clip]{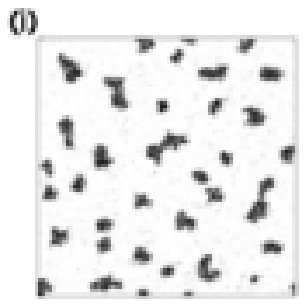}}
 {\includegraphics[width=0.24\linewidth,clip]{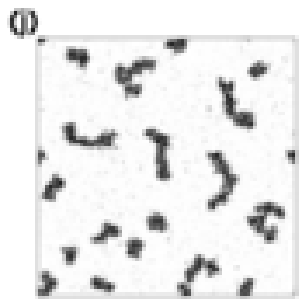}}
 {\includegraphics[width=0.24\linewidth,clip]{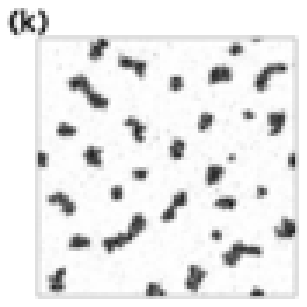}}
 {\includegraphics[width=0.24\linewidth,clip]{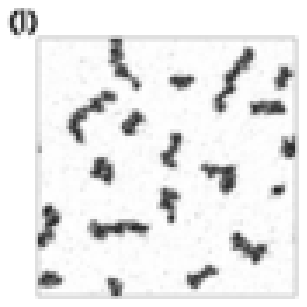}}
 {}
 \caption{}
 \label{lambda_snapshots}
\end{figure}

Figure \ref{lambda_snapshots}:
{Snapshots of the dynamics simulations for amphiphilic diblock copolymer
solutions with various cutoff length $\lambda$. Black color represents $\phi_{B}(\bm{r})$.
(a),(b) $\lambda = 1$, $\tilde{t} = 312.5, 1250$.
(c),(d) $\lambda = 2$, $\tilde{t} = 312.5, 1250$.
(e),(f) $\lambda = 4$, $\tilde{t} = 312.5, 1250$.
(g),(h) $\lambda = 5$, $\tilde{t} = 312.5, 1250$.
(i),(j) $\lambda = 6$, $\tilde{t} = 312.5, 1250$.
(k),(l) $\lambda = 8$, $\tilde{t} = 312.5, 1250$.
}

\clearpage

\clearpage

\clearpage

\clearpage

\clearpage

\clearpage

\clearpage

\clearpage

\clearpage

%------------------------------------------------------------------------------
\end{document}